\documentclass[]{aa}
\usepackage{natbib}
\usepackage{graphicx}
\usepackage{wasysym}
\usepackage{txfonts}
\usepackage{color}

\def\HH{\mbox{H$_2$}}
\def\fH2{\mbox{f$_{{\rm H}_2}$}}

\def\EBV{\mbox{E$_{\rm B-V}$}}

\def\AV{\mbox{A$_{\rm V}$}}

\def\nH2{\mbox{${\rm n}_\HH}$}

\def\pccc{~{\rm cm}^{-3}} 
 
\def\pcc{~{\rm cm}^{-2}}

\def\Tsub#1 {\mbox{${\rm T}_{\rm #1}$}}
\def\TK  {\Tsub K }

\def\arcsec{\mbox{$^{\prime\prime}$}}

\def\degr{$^{\rm o}$}
\def\p{\mbox{$^+$}}
\def\m1{\mbox{$^{-1}$}}

\def\hocp{\mbox{{HOC\p}}}

\def\cch{\mbox{C$_2$H}}
  
\def\hhco{\mbox{H$_2$CO}}
\def\h13cop{\mbox{{H$^{13}$CO\p}}}

\def\C3H{\mbox{C$_3$H}}
\def\c3h2{\mbox{C$_3$H$_2$}}
\def\cc3h2{\mbox{{\it c}-C$_3$H$_2$}}
 
 \def\R0{R$_0$}
\def\G0{\mbox{G$_0$}} 
  
  \def\deg{{}^\circ}
\def\ddeg{{}^\circ\kern-.1em}

\def \kms{\mbox{km\,s$^{-1}$}}

\def\bll{BL Lac}

\def\E#1 {$10^{#1}$}
\def\E#1 {E{#1}}
\def\P#1,{$\nH2\TK~=~#1\times~10^4\pccc$~K}
\def\ec#1,#2,#3,{#1\,(#2)\E{#3}}

\def\H3{\mbox{H$_3$}}

\def\RH2{\mbox{R$_{\rm G}$}}
\def\g13{\mbox{g$_{13}$}} 

\def\cc3h{\mbox{{\it c}-\C3H}}
\def\lc3h{\mbox{{\it l}-\C3H}}

\def\Whcop{\mbox{$\Upsilon_{{\rm HCO}\p}$}}
\def\Wcch{\mbox{$\Upsilon_{{\rm CCH}}$}}
\def\Whcn{\mbox{$\Upsilon_{{\rm HCN}}$}}

\def\L21{\mbox{{$\lambda$21cm}}}

\newcommand{\emm}[1]{\ensuremath{#1}}   
\newcommand{\emr}[1]{\emm{\mathrm{#1}}} 

\newcommand{\hcop}{\emr{HCO^+}} 
 
\newcommand{\cotw}{\emr{^{12}CO}}
\renewcommand{\coth}{\emr{^{13}CO}}

\newcommand{\X}[1]{\emm{X_\emr{#1}}}

\newcommand{\XCO}{\X{CO}}

\newcommand{\W}[1]{\emm{{\rm W}_\emr{#1}}}
\newcommand{\WCO}{\W{CO}}

\newcommand{\Msun}{\emm{M_\odot}}

\def\nix#1 {}

\title{CO, CS, HCO, HCO$^+$, C$_2$H, and HCN in the diffuse interstellar medium}

\author{H. Liszt\inst{1} and M. Gerin\inst{2}}
\institute{
     National Radio Astronomy Observatory,
           520 Edgemont Road,
           Charlottesville, VA,
           USA 22903 
      \email{hliszt@nrao.edu}
\and
LERMA, Observatoire de Paris, PSL Research University CNRS Sorbonne Universit\'e \\
\email{maryvonne.gerin@observatoiredeparis.psl.eu}
}

\begin{document}
\date{received \today}
\offprints{H. S. Liszt}
\mail{hliszt@nrao.edu}

\abstract
{Radio frequency molecular absorption lines appear 
 along sight lines with \AV\ well below 1 mag, revealing  the
 presence of H$_2$ in  diffuse gas even when $\lambda$2.6mm CO emission is absent.} 
{We discuss absorption lines of \hcop, \cch, HCN, CS, and
HCO in a larger sample (88 sight lines) than was available before.} 
{We observed millimeter-wave absorption at the Institut de radioastronomie millimetrique (IRAM) and 
Atacama Large Millimeter/submillimeter (ALMA) interferometers 
over the  past 30 years and gathered the results for to compare
with observations of \hcop\ and CO emission taken at the Arizona Radio Observatory (ARO) KP12m and 
IRAM 30m telescopes.} 
{We detected \hcop\ along 72 of 86 sight lines where it was observed, 
\cch\ along 53 of 76 sight lines, and HCN along 38 of 57 sight lines. 
\cch\ is ubiquitous, and N(\cch)/N(\hcop) increases at smaller \EBV\ and smaller N(\hcop), 
but \cch\ absorption is intrinsically weaker, which lowers the number 
of sight lines with a low column density along which it was detected.
The dense-gas tracer HCN 
was uniformly detected down to N(\HH) $ = 10^{20}\pcc$, with little change in 
the relative abundance of N(HCN)/N(\hcop) $\approx 1.25$. HCO was detected along 
only 4 of 46 sight lines that were newly observed at ALMA because its lines are intrinsically weak, but HCO is ubiquitous in the interstellar medium with N(HCO)/N(\hcop) = 1/3 or 
N(HCO)/N(\HH) $= 10^{-9}$.  CS was not seen in the 13
directions where it was newly observed, which are all in Chamaeleon.  The line widths 
of HCN features are (like those of CO) narrower than those of matching features 
observed in \hcop, and those of \cch\ are broader.  \hcop\ emission is commonly
observed at (log) levels $-2\pm0.3$ dex with respect to CO emission. 
The comparison of \hcop\ emission and absorption along 7 sight lines yielded 
a typical \hcop\ excitation temperature of 2.76 K and total neutral hydrogen number 
densities n(H) of $\approx 50-200 \pccc$.}
{}

\keywords{Astrochemistry; Galaxy: disk; ISM: abundances; ISM: atoms; ISM: clouds; ISM: dust, extinction; 
ISM: molecules; Line: profiles;Molecular processes}

\authorrunning{L\&G} \titlerunning{DNM, dark and molecular gas}

\maketitle{}

\section{Introduction}

In several  recent papers \citep{LisGer+18,LisGer+19,LisGer23b}, we used the Atacama Large Millimeter/Submillimeter Array (ALMA) 
observations of millimeter-wave \hcop\ and CO absorption to demonstrate the molecular character 
of the dark neutral medium (DNM) that is present at the HI-\HH\ transition in 
 diffuse or translucent interstellar gas \citep{GreCas+05,Pla15Cham,RemGre+17,RemGre+18}.
Gas in the DNM is inferred to be present from far-infrared (FIR) dust emission
and $\gamma$-ray emission 
but is imperfectly traced and largely invisible in $\lambda 21$cm HI and 
$\lambda 2.6$mm CO emission. In our work,  sight lines with a low CO column density
that lack detectable CO emission showed 
\hcop\ absorption when the reddening exceeded the known threshold 
\EBV\ $\ga 0.08$ mag \citep{SavDra+77,LisGer23a} for observing \HH\ fractions 
\fH2\ = 2N(\HH)/(N(HI)+2N(\HH)) = 2N(\HH)/N(H)$\ga\ 0.1$.  With a few exceptions that
imply a possible saturation of the $\lambda 21$cm HI profile, the newly inferred 
\HH\ column densities N(\HH) = N(\hcop)/$3\times10^{-9}$ closely matched the amount 
of DNM.

In the course of these observations along 13 sight lines in Chamaeleon and 33 
sight lines in the Galactic anticenter, we also used the prodigious 
data-gathering abilities of ALMA to observe absorption from other species that were previously 
detected in diffuse gas, 
\cch\ \citep{LucLis00C2H,LisSon+12}, HCN \citep{LisLuc01}, CS \citep{LucLis02}, 
HCO \citep{LisPet+14}, and \hhco\ \citep{Nas90,LucLis96,LisLuc+06}. We discuss these associated observations here. In combination
with ALMA observations of 4 sight lines at low latitude in the inner Galaxy
\citep{GerLis17,LisGer18} and the previously published absorption data from 
the IRAM 30m telescope and Plateau de Bure Interferometer (PdBI), we discuss observations along
86 sight lines in the most commonly observed species \hcop, 76 in \cch\,
and 57 in HCN. HCO is a ubiquitous species with weak absorption 
\citep{LisPet+14}
that was detected along all 4 of the 46 sight lines at which it was expected based on the  ac achieved rms noise levels.

Far fewer (25) sight lines were observed in total in CS after our failure to 
detect CS absorption along the 13 sight lines in Chamaeleon within the 
integration times that sufficed to detect \hcop\ and \cch. No new detections
of CS are presented here, and, as discussed below, the new upper limits on CS are not 
significant because the molecular hydrogen column densities that characterize
the observed sight lines are low.
A comparison of absorption line observations of CS and \HH CO toward B0355+508 
recently showed that the relative abundance of CS increases in presumably denser
gas with a smaller electron fraction, where a substantial fraction of ionized 
carbon has recombined to neutral atomic carbon \citep{GerLis+24}. This suggests a 
role of neutral carbon in the formation of CS along with recombination of HCS\p\ that 
was also inferred \citep{LucLis02}.  CS J=2-1 emission from diffuse gas is bimodal 
with respect to \hcop\ emission and abruptly increases in brightness when the 
integrated brightness of CO exceeds 10 K-\kms\ \citep{Lis20}.

Section 2 summarizes the observational material we considered. Section
3 discusses the properties of the sample in the broader Galactic context
and summarizes the abundance relations among \hcop, \cch, HCN,  CS, and HCO.
Section 4 compares the line widths of Gaussian kinematic components fit to \hcop, 
\cch\, and HCN. In Section 5 we use the \hcop\ and CO emission to derive \hcop\ 
excitation temperatures and hydrogen number densities in the host gas and 
to relate the \HH\ sampled in \hcop\ absorption to  the 
CO emission brightness,  the \cotw/\coth\ brightness ratio, and the CO-\HH\ conversion
factor. These quantities are commonly used to characterize the gas. Section 6 presents our summary.

\section{Observations and data sources}

The dataset we discuss here is a combination of  older and newer observations,
as noted in Table \ref{TableDataSources}. The 
older absorption line work is from the IRAM PdBI and 
NOEMA interferometers, and the older CO and \hcop\ emission line profiles were taken at the previous-generation 
ARO 12m telescope on Kitt Peak.  The newer observations of the absorption line were taken 
at ALMA, and the newer CO and \hcop\ emission profiles were taken at the IRAM 30m 
telescope. The older data and some of the newer data have been described in earlier
work cited here.  The portions of the current dataset that have not been published
previously are the 87-98 GHz ALMA absorption profiles of HCO, HCN, \cch\, and CS, 
and the profiles of \coth\ and \hcop\ emission that were taken in the course
of our recent work to characterize the DNM, as noted in the Introduction.

\subsection{Absorption line spectra}
As in our earlier work, the new ALMA absorption line profiles  were 
taken from the continuum-peak pixels of the spectral line 
datacubes from the standard pipeline product for the various species, and they were normalized by the continuum flux in the 
appropriate continuum map of the spectral window.  The channel spacing and approximate 
spectral resolution of the ALMA data were 30.5 and 61.0 kHz, or 0.102 and 0.205 
\kms\ respectively, at the \hcop\ rest frequency 89.1885 GHz.  All directions
in the Galactic anticenter were observed for the same time, which resulted in strongly 
varying rms line/continuum sensitivities.  

We newly observed the F=2-1 and F=1-0 hyperfine components of the N=1-0, J=3/2-1/2 
transition of \cch\, whose strongest hyperfine component (F=2-1) is at 87.317 GHz. 
The observed components included  5/8 of the total integrated 
line strength of the N=1-0 J=3/2-1/2 transition. We define the total as \Wcch.
N(\cch) $= 2.71 \times 10^{13} \pcc$\Wcch\ as in our earlier survey
of \cch\ absorption \citep{LucLis00C2H}, where  all hyperfine components were observed.
The spectra of \cch\ are shown in Figure \ref{FigureC2H}.

As in earlier work \citep{LisLuc01}, we observed the J=1-0 transition
of HCN, whose strongest hyperfine component is at 88.632 GHz, and we 
derived N(HCN) $ = 1.89 \times 10^{12} \pcc$ \Whcn\, where \Whcn\ 
is the integrated optical depth summed over all three hyperfine components.
The spectra of HCN are shown in Figure \ref{FigureHCN}.

The line-of-sight properties and measured absorption line quantities are given in Table \ref{TableLOS}
for \hcop, HCN, and \cch.  The fluxes in this table are averages of the almost equal 
fluxes in the three spectral windows. Upper limits are the 
$3\sigma$ values
measured over the span of the \hcop\ line, when given. In the five cases when \hcop\ was
not observed or not detected, they are given over an arbitrary 3 \kms\ interval. 
No absorption from CS was reliably detected in our observations of Chamaeleon. We therefore did not observe CS in the Galactic anticenter.

\subsection{Emission line spectra}

The IRAM 30m emission profiles have a channel spacing and approximate spectral resolution of 48.8 and 97.7 kHz, or 0.164 and 0.33 \kms\ at the \hcop\ rest frequency. 
 
The \hcop\ profiles we show are averages of spectra displaced by 1.2 HPBW = 30\arcsec\ from the continuum source in each of the four cardinal directions, as discussed in Section \ref{SectionA.5}. This was necessary because the emitted and absorbed fluxes are comparable, and no pure emission spectrum was obtained toward the continuum at the 30m telescope. This is not a concern for much brighter CO lines.    
 
As before, the emission results are presented on the main-beam antenna temperature scale that is native to the 30m telescope.

\begin{figure*}
\includegraphics[height=15.5cm]{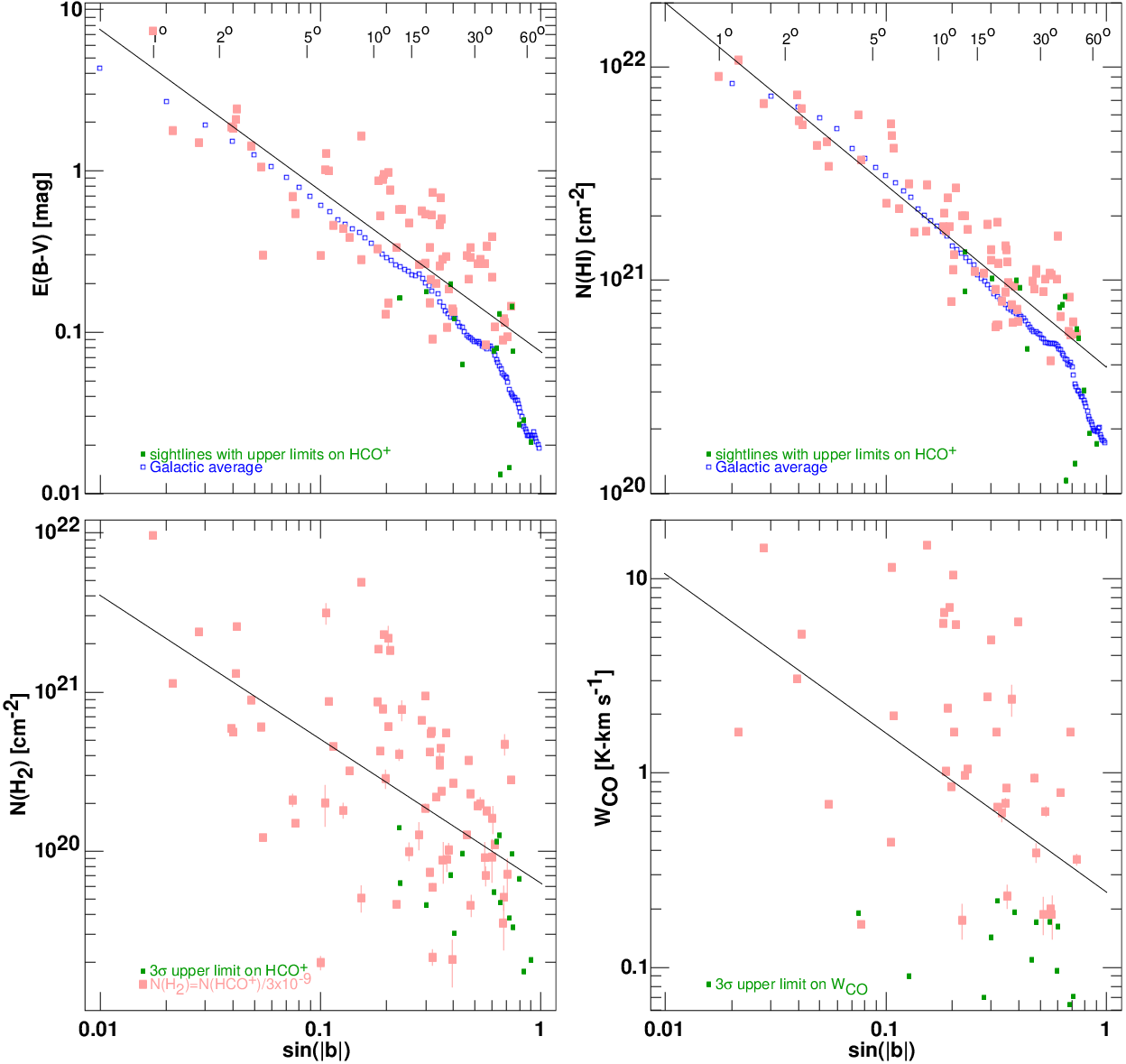}
\caption[]{Galactic-scale properties of the sample. We plot against
$\sin(|b|)$ \EBV\ (top left), N(HI) (top right), 
N(\HH) = N(\hcop)/$3\times 10^{-9}$ (bottom left), and \WCO, the line 
profile integral of the CO J=1-0 emission. The sight lines that were not detected in \hcop\ absorption are shown as smaller green symbols
in the upper two panels.  The smaller green symbols in the lower panels are the $3\sigma$ 
upper limits on the plotted quantity. The solid line in each panel is a
regression line: A plane-parallel vertically stratified Galactic gas layer would 
exhibit a power-law slope of -1. We also show averages
at intervals of 0.01 in sin($|b|$) in blue at the top. A legend marking the values of the Galactic
latitude is inset near the top of the upper two panels. The values for the statistical parameters 
and the regression fits are given in Table 1.}

\end{figure*}

\subsection{Atomic and molecular hydrogen column densities}

The $\lambda 21$cm HI column densities were derived from the Leiden, Argentine, and Bonn
(LAB) All-Sky HI survey \cite{KalBur+05} in the optically thin limit as
N(HI) = $1.823\times10^{18}\pcc$ 
W$_{\rm HI}$, where W$_{\rm HI}$ is the $\lambda21$cm HI emission line profile integral in units of K-\kms. 

As is common practice (see \cite{GerLis+19}), we derived the 
\HH\ column densities as N(\HH) = N(\hcop)/$3\times10^{-9}$
and N(\hcop) $= 1.10\times10^{12} \pcc$ \Whcop\, where \Whcop\ is
the integrated optical depth of the J=1-0 \hcop\ absorption line,
which corresponds to a permanent dipole moment of $\mu =3.89$ Debye for \hcop.
Much of our earlier work (eg \cite{LucLis96}) employed a higher permanent dipole moment of $\mu =4.07$ Debye and derived $\approx 10$\% lower N(\hcop) for a given \Whcop.

\subsection{Conventions}

The velocities we presen with the spectra were taken with respect to the kinematic 
definition of the local standard of rest.  We defined N(H) as the column 
density of H nuclei detected in their neutral atomic and molecular form, 
N(H) = N(HI)+2N(\HH). The molecular hydrogen fraction \fH2\ =  2N(\HH)/N(H).
The integrated absorption of the \hcop\ profile in units of 
\kms\ is denoted by \Whcop\ and similarly for other species.  The integrated 
emission of the J=1-0 CO line is denoted by \WCO\ in units of K-\kms\ ,
and the CO-\HH\ conversion factor N(\HH)/\WCO\ is denoted by \XCO.

Where reference is made to a typical Galactic or standard CO-\HH\ conversion 
factor, the value N(\HH)/\WCO\ $= 2\times10^{20}~\pcc$/(1 K-\kms) should
be understood, as summarized in Table E.1 of \cite{RemGre+17}. Where we refer 
to optical reddening, the values cited were taken from the work of \cite{SchFin+98}.

Where the contributions of individual datasets are noted in the figures, the 
rubric PdBI denotes results taken from the earlier surveys of 
\hcop\ \citep{LucLis96}, \cch\  \citep{LucLis00C2H}, HCN \cite{LisLuc01}, and 
CS \citep{LucLis02}. The rubric Chamaeleon refers to  new data taken
incidental to the DNM studies of \cite{LisGer+18,LisGer+19} ,
and Inner Galaxy refers to the data of \cite{GerLis17} and \cite{LisGer18}.
New results incidental to DNM studies in the Galactic anticenter \citep{LisGer23b}
use the rubric Outer Galaxy.  

\subsection{Distances and velocity assignments}

Approximate cloud distances derived from the GAIA stellar reddening surveys are 
available for sight lines in or near well-known cloud complexes that have been 
specifically targeted, for instance, 180 - 200 pc in Chamaeleon 
\citep{VoiJor+18,YanQin+19,ZucSpe+19} and 140 pc in Taurus \citep{YanQin+19,ZucSpe+19}.  
There are two isolated sight lines \citep{LisPet12,Lis20} in the general vicinity of 
the high-latitude cloud MBM 53 at 250 pc \citep{SunMin+21}, and the gas toward \bll\ is 
a few degrees away from the Lacerta region identified at 500 pc by \cite{YanQin+19,ZucSpe+19}. 

The 3D extinction maps of \cite{LalVer+22} show the run of
extinction with distance along our specific sight lines. The relatively large jumps 
in extinction associated with molecular gas uniquely associate dust structures 
with the absorption line velocity in the simpler cases, even in the absence of CO 
emission.  Conversely, structure in the run of extinction with distance may help us to  understand kinematic substructure in the absorption line profiles. This subject
will be discussed in a future contribution.  
        
\section{The sample}

The sample is an amalgamation of datasets that were taken over the course of nearly 30 years for various 
purposes, at instruments that overlap only partially in their sky coverage.  The PdBI surveys of 
\hcop\ \citep{LucLis96} and \cch\ \citep{LucLis00C2H} contained flux-limited samples that were augmented 
with a few sources of particular interest, as viewed from the perspective of the Northern Hemisphere, and the 
PdBI data for HCN and CS were much more limited.  
The ALMA data in Chamaeleon were complete in flux over a very particular sky region.  The newer ALMA
data in the outer Galaxy were derived from a serendipitous sample of sight lines that were previously
studied to derive their DNM content. The properties and biases of the sample are discussed in this
section.

\subsection{Galactic-scale properties}

Figure 1 shows the galactic-scale properties of our sample. The gas 
and dust column densities are plotted against  $\sin(|b|)$, which is proportional to the nominal path length through a plane-parallel 
stratified  Galactic disk. The numerical values derived from these plots are given 
in Table 1. In the top left corner in Figure 1, the reddening \EBV\ along 
individual sight lines \citep{SchFin+98} is plotted, together with the all-sky 
average at each latitude calculated at increments of $\Delta\sin(|b|) = 0.01$. The 
regression line for the sample has a power-law slope of
$-1.00\pm0.08$, which is appropriate for a plane-parallel stratified medium, but the
Galaxy per se progressively departs from this geometry for $|b| \ga 20\deg$. The 
sample values of \EBV\ lie above the Galactic average at all latitudes and do not 
show the same steep decline until $|b|\ga 44$\degr.  Thus, the sample is noticeably 
rich in dust and gas. We also mark the sight lines without \hcop\ detections.
These sight lines track the Galactic average more closely and mostly lie below the
regression line. 

Figure 1 in the top right corner shows a similar plot for the HI profile integral converted
into column density N(HI), as described in Section 2.3. In this plot, the sample 
mean N(HI) follows the Galactic average for $|b| \la 15\deg$ and 
departs less from the Galactic average up to about $30\deg$ than \EBV.  
When the departure of 
the sample from the Galactic average is smaller in N(HI) than in \EBV, the 
sample is likely to be rich in molecular gas. In contrast to the behavior of 
\EBV, the sample sight lines without a detection of \hcop\ have typical sample 
values of N(HI) up to very high latitudes, where the sample values of \EBV\ 
decrease. It is tempting to identify the missing \EBV\ with the 
absent \HH.

Figure 1 shows the Galactic-scale distribution of two quantities
measured in this study in the bottom panel, N(\HH) = N(\hcop)$/3\times10^{-9}$ at the left (where the 
upper limits on \hcop\ were included in the calculation of the regression line 
at their displayed values), and \WCO\ at the right. The \hcop\ column 
density and CO emission brightness show more scatter owing to the sensitivity 
of the \HH\ and CO formation to local gas properties. The sight lines without a \hcop\ 
detections fall well below
the regression line at the left, but are well mixed in the general data population.  
In any case, \hcop\ was not detected along any sight line above $44\deg$ ,
corresponding to $\sin(44\deg)=0.7$.

Figure 1 shows in the bottom right panel the data for the line profile integral of CO 
emission \WCO.  In this plot,  the upper limits on CO emission fall far below 
the weakest detections. This differs from the plot in the bottom left panel, where the 
upper limits on the \hcop\ absorption are at levels at which \hcop\ is commonly 
detected. We generally failed to observe any CO emission at levels 
\WCO\ $\la  0.2$ K-\kms.  No CO emission was 
detected along sight lines without a detection of \hcop. The sight lines without CO emission when \hcop\ absorption is detected are possibly more interesting.

The results in Figure 1 indirectly show that the molecular fraction of the diffuse 
gas is substantial (1/4-1/3), the applicable CO-\HH\ conversion factor is near 
the standard Galactic value (Section 2.4), and even very low levels of CO emission 
can imply significant molecular fractions in the ISM.

The HI column through the Galactic disk vertically at the solar radius is 
usually quoted as N(HI) $\approx 3\times10^{20}\pcc$, which is slightly below the 
90\degr\ intercept of the  regression line for N(HI) in the upper right panel in 
Figure 1, $\approx 4\times10^{20}\pcc$. By comparison, the 90\degr\ 
intercept of the regression line for N(\HH) in the lower left panel is 
N(\HH) $\approx 0.6\times10^{20}\pcc$, implying 
\fH2\ = 2N(\HH)/(N(HI)+2N(\HH)) $\approx$ 1/4-1/3.

The vertical intercept of the regression line in the lower left panel is at \WCO\ = 0.245 K-\kms,
compared to \WCO\ = 0.235 K-\kms\ derived from the smaller sample of older data 
cited here \citep{LisPet+10}.  In comparison with the result in the lower left panel 
for \HH, the implied CO-\HH\ conversion factor is 
$0.63\times10^{20}\pcc/(0.245$K-\kms) $= 2.57\times10^{20}~\pcc$/(K-\kms).

If we were to look at the Milky Way vertically from outside the solar circle, the 
CO brightness would be $2\times$0.245 K-\kms\ = 0.49 K-\kms. This is small
compared to the brightness of disks in external galaxies such as M51 
\citep{BigLer+16}, but it is about half of the face-on brightness
of the Milky Way at the solar circle, 0.8 K-\kms\ \citep{LisPet+10}.
The diffuse molecular gas causes about half of the molecular 
hydrogen in the ISM near the Sun. An integrated brightness of 1 
K-\kms\ is equivalent to a \HH\ mass of 4.1\Msun\ pc$^{-2}$ using the CO-\HH\ conversion 
factor derived above.

\subsection{Molecules mixed in the Galactic gas and dust}

Figure 2 plots the column densities of \hcop, \cch, HCN, and CS against reddening 
\EBV. \hcop\ is the most commonly detected molecule considered here, but neither 
it nor any other molecules were found at \EBV $< 0.07$ mag where the molecular 
hydrogen fraction \fH2\ is generally small in the ISM \citep{BohSav+78}. The precise 
\EBV\ locus of the onset of appreciable values of \fH2\ is somewhat uncertain 
\citep{LisGer23a} . 

Table 2 shows the fractions of detections of \hcop, \cch, HCN, CS, and HCO
with \EBV\ (in  decreasing order of the number of detections) and with the
choice of a $2\sigma$ or $3\sigma$ confidence level. \EBV\ = 0.0675 mag is the 
lowest reddening at which any absorption was detected in \hcop.
\hcop\ is detected 
two-thirds of the time at \EBV\ $\le 0.125$ mag or \AV\ $\le 0.4$ mag, and 
the detection rates of \hcop\ are almost entirely unaffected by the choice 
of confidence level. The ubiquity of \hcop\ absorption enhances 
the ability of detecting other species by providing a guide as to
which sight lines harbor molecular gas and the velocity range over
which molecular absorption or emission in other species is expected.

\cch\ is the second most commonly detected molecule. Its lower detection 
rate (compared to \hcop) results more from a lower sensitivity than from a
narrower, less ubiquitous distribution in the Galactic gas.  Overall,
\cch\ was detected in two-thirds of the directions in which it was observed.

The so-called high-density tracer HCN is detected at least half the time 
when \EBV\ $\ga 0.125$ mag or \AV $\ga 0.4$ mag and it is detected along two-thirds 
of the sight lines where it was observed. This is almost as frequently 
as \cch.  CS, another high-density tracer, was the least often detected 
species of the four species shown in Figure 2 for reasons that are 
discussed in Section 3.4. 

The numerical values of statistical properties and power-law regression fits 
derived from the plots in Figure 2 are given in Table 2. \hcop\ has a power-law 
slope versus \EBV\ of $1.03\pm0.11$ in Figure 2, but the Galactic anticenter 
sample shows a very rapid increase in \hcop\ with \EBV\ within the overall envelope 
of the data. The mixing of gas and dust over long sight lines with high reddening near the 
Galactic plane in the inner-Galaxy sample tends to drive the overall behavior 
toward the unit slope, while the abundances are more likely to be dominated 
by localized cloud-level behavior when the \EBV\ is modest. 

HCN alone shows a power-law slope higher than one overall,  $1.35\pm0.10$. The slope of the 
power-law fit to N(\cch), 
$0.83\pm0.10$, is noticeably shallower than that of \hcop. This indicates an increase 
in N(\cch)/N(\hcop) at smaller \EBV\ that was reported earlier \citep{LucLis00C2H}. 
The plot in the lower right panel shows  that CS is lower at \EBV\ $\la 0.6$ mag
than in the regression line for the detected lines of sight.  

The abundances are broadly consistent with earlier results, 
but they are subject to several influences, including modern use of a smaller permanent 
dipole moment for \hcop, 3.89 versus 4.07 Debye.  This leads to an  increase of 9.4\% in 
N(\hcop) and to a smaller abundances relative to \hcop\ and \HH\ for other species.
Where the abundances have a nonlinear relation, the derived average relative abundances
depend on the makeup of the samples, as discussed in Section 3.4. 

\subsection{HCO}

This work and the recent detection of HCO toward J1851+0035 by \cite{NarKan+24} 
provided the first observations of HCO in diffuse gas since the original detection 
by \cite{LisPet+14}. HCO is a widely distributed species whose lowest rotational 
transition is spread over four hyperfine components with LTE line strengths in a 4:2:2:1 
ratio. The strongest hyperfine component at 86670.76 MHz was observed to be 2\% as 
strong as \hcop\ \citep{LisPet+14}, meaning that HCO can only be detected at the 
$2\sigma$  level in directions where \hcop\ is detected with a signal-to-noise ratio of 100 
or more. There were 4 
directions like this among the 46 observations in which HCO was observed in this work, and HCO was 
detected above the $2\sigma$ level with a relative abundance 
N(HCO)/N(\HH) $= 1.1-1.3\times 10^{-9}$ along four sight lines, as detailed in 
Table 5. As noted there, HCO was previously detected toward \bll\ at
\EBV\ = 0.32 mag with a relative abundance $= 0.98\pm0.02 \times 10^{-9}$ based 
on the detection of all four hyperfine components. HCO is about one-third as abundant 
as \hcop\ and 25 times more abundant than \hocp\ \citep{LisLucBla04,GerLis+19}.

\subsection{Inter-relations in molecular species}

Figure 3 plots the variation in \cch, HCN, and CS against \hcop in the left panel and 
the variation in \hcop, HCN, and CS against \cch in the right panel. The numerical values of 
statistical properties derived from these plots are given in Table 5. 
The slopes of the regression lines in the plots versus N(\cch) in the right panel are all
steeper than those in the left panel, 1.01 versus 0.67, 1.40 versus 1.25, and 1.4 versus 1.1, 
from top to bottom.

The slope of the regression line fit of N(\cch) versus N(\hcop) in the top left panel is
0.67, which confirms the higher relative abundance ratio N(\cch)/N(\hcop)
at smaller N(\hcop) seen in a substantial but smaller and more homogeneous
earlier survey of high-latitude northern targets \citep{LucLis00C2H}. 
This nonlinearity means that the mean \cch/\hcop\ ratio varies with 
the composition of the sample, and the current mean, 
$<$N(\cch)$>/<$N(\hcop)$>$ = 10.6, is slightly lower than the value 
$<$N(\cch)/N(\hcop)$>$ = 14.5 quoted earlier. 

HCN is detected down to N(\hcop) $\la 4\times10^{11}\pcc$ with a tight 
correlation (r = 0.95) and without a break in the power-law slope ($1.25\pm0.06$) 
of the HCN-\hcop\ relation.  Earlier, sparser data (Figure 3 of \cite{LucLis02}) 
seemed to show an abrupt increase in N(HCN) at N(\hcop) $\approx 10^{12}\pcc$.
The HCN/\hcop\ abundance ratio was quoted earlier as 0.7 - 2.5 
(their Table 1).  The current
sample mean is $<$N(HCN)$>/<$N(\hcop)$> = 1.25\pm0.06$, and a comparison
of the Gaussian decomposition products gives a slightly higher number (Sect. 4),
but the regression line in Fig. 3 is characterized by values below unity
for N(\hcop) $\la 10^{13}\pcc$.  

The upper limits on N(CS) in Chameleon and the values derived  earlier 
by \cite{LucLis02} 
are statistically significant in showing a deficiency at \EBV\ $\la$ 0.5 mag 
relative to the regression line for the  detections at higher extinction
in Fig. 2. They are, however,
too high to provide useful information for the CS chemistry of the 
sight lines at \EBV\ $\la 0.5$ mag given the plot in the lower right panel
in Fig. 3.  The integration times needed to study the other molecules 
were in general too brief to provide much useful information on CS
beyond that noted in the Introduction.

\section{Component decomposition}

Figures 4 and 5 compare the results of a Gaussian decomposition of the 
new profiles of \hcop, HCN, and \cch. For each species, the decomposition 
simultaneously fit all kinematic components and hyperfine components by
varying the central optical depth, 
central velocity, and profile FWHM while minimizing deviations from the observed 
line/continuum ratio $\exp(-\tau)$. In this way, the quoted rms error estimates 
appropriately reflect the increased errors encountered when the line/continuum 
ratios were fit at high optical depth in a spectrum with a fixed rms flux/continuum noise 
level, and the uncertainty introduced by multiple components.

We used the results for \hcop\ as an initial template for HCN and \cch\ in
decompositions that considered  all observed kinematic and hyperfine components. 
The hyperfine components of HCN and \cch\ were fit assuming optical depth ratios 
proportional to their intrinsic line strengths.  The associated numerical results are 
given in Tables 6 and 7.

The results considered matched and unmatched components. The matched components
are those for which the separation in central velocity is smaller than one-third
of the FWHM of the narrower of the possible pair of components. Ambiguities
are introduced in the case of weak broader components, and the component list
was therefore ruthlessly purged to achieve a minimum set of matched lines. As noted
below, the matched sets account for the vast majority of the aggregate optical
depth in every species, including \hcop\,whose profiles accommodate a much
larger number of fitted components.

\subsection{Line widths}

The Gaussian components fit to CO absorption were unambiguously found to be 
10-15\% narrower than those of \hcop\ along six sight lines in Chamaeleon 
\citep{LisGer+19}.  The results for line width comparisons between \hcop\ and
HCN or \cch\ find small differences with a lower significance.  In any case, the
new results did not deviate from a linear proportionality of the 
derived FWHM (Table 6 and Table 7).

The typical line widths of the matched components for all species observed here are very 
similar, 1 \kms\, while the matched HCN components are found to be 5\% 
narrower on average than in \hcop\ (upper left panel in Figure 4 and Table 6). As shown in Table 6, 
\hcop\ was decomposed into many more components than HCN along the 28 sight lines 
decomposed in \hcop\ and HCN (107 versus 55), but the 52 unmatched components in \hcop\ 
represent only 12\% of the aggregate \hcop\ absorption, and the four unmatched components 
of HCN represent 2\% of the HCN absorption.  The unmatched \hcop\ components are 
much weaker and somewhat broader than average, as is the case for the unmatched 
HCN components. 

Absorption from \cch\ is intrinsically weaker than in \hcop\ or HCN, and somewhat different 
results were obtained when we compared \hcop\ and \cch\ along 31 
sight lines that were decomposed in both species. For \cch, 73 of 119 \hcop\ components 
were matched with 73 of 81 components fit to \cch\ (Figure 4 in the bottom panel and Table 7),
and \cch\ is 12\% broader, with considerably more scatter than for the comparison
of \hcop\ and HCN.

\hcop\ is more easily detected owing to its higher dipole moment and lack 
of hyperfine structure, and the many weak broad unmatched \hcop\ 
components result in a cumulative distribution function for \hcop\ that is shifted to 
higher line widths in Figure 5 in the left panel, where all components that we fit to each species were 
used to construct the cumulative distribution function, not only the components
that were matched in two species.

\subsection{Component column densities}

The component column densities are shown in the bottom panel in Figure 4 and are summarized 
quantitatively in Tables 6 and 7. The fits mimic the behavior seen in Figure 
3 and Table 5, with power-law slopes $\approx$ 1.2 and 0.7 for HCN and \cch\ ,
respectively.  More \hcop\ components were matched in \cch\ than in HCN 
(73 of 119 versus 55 of 107), with many matched components 
at N(\hcop) $< 10^{12}\pcc$. The mean \hcop\ column density of components
matched in HCN, $1.26 \times 10^{12}\pcc$, is larger by one-third than that
for components matched in \cch\ ($0.92 \times 10^{12}\pcc$ ; Table 7). The relative abundance ratio 
of the weakest matched components N(\cch)/N(\hcop) $\approx 100$  
exceeds the overall mean $<$N(\cch)$>/<$N(\hcop)$> \approx 25$ by far.
 
\section{HCO$^+$ and CO in absorption and emission}

Seven of the Galactic anticenter sight lines were newly observed in \coth\ and \hcop\ emission, as shown in Figure 6.  The comparison of \hcop\ observed in emission and absorption can be used to determine the excitation temperature of the J=1-0 
transition of \hcop\ and the volume density of the host gas. The comparisons of \hcop, \cotw\, and \coth\ in various combinations of emission and absorption
reveal information on the overall character of the gas as seen in 
CO: the CO brightness, \cotw/\coth\ ratio, and CO-\HH\ conversion factor
at which the molecular gas is found. 

The results are presented in two formats. In one format, ALMA and IRAM 30m profiles 
were interpolated onto common spectral grids for a channel-to-channel comparison 
of the new results between species and tracers. In the other format, the results 
integrated across whole profiles are shown for the larger sample that includes
older data.

\subsection{HCO$^+$ compared in absorption and emission}

The left panel of Figure 7 shows the brightness of the \hcop\ emission plotted against the 
\hcop\ optical depth on a channel-by-channel basis.  Emission is detected in a 
statistical sense at optical depths $\tau \ga 0.05$.  The emission toward J0502 is 
much brighter than in any other source and accounts for almost 50\% of all emission.
For most sources, the excitation temperature inferred from the regression
line passing through 0.015 K emission brightness at $\tau = 1$ is 2.758 K.
For J0502, the excitation temperature corresponding to a brightness of 0.09 K
at $\tau = 2$ is 2.855 K. 

The right panel of Figure 7 plots the integrated \hcop\ emission brightness and optical depth 
for the new and old results together with iso-density lines derived from the weak 
excitation limit (including electrons) in the discussion of the detectability of millimeter-wave 
emission by \cite{LisPet16}. 
 When the collisional excitation of a species is sufficiently far below thermalization, all energy deposited in the rotation ladder by collisions escapes the medium, regardless of the optical depth. The emission line brightness is determined by the product of the column density of the molecule, the mean hydrogen number density, and an effective excitation rate coefficient that includes all  upward collisional excitations into whatever energy levels. The rate coefficient is 30-100 times larger under diffuse gas conditions where the ionization fraction exceeds $10^{-4}$
compared to fully molecular gas, where CO dominates the carbon budget and the ionization fraction is $10^{-7}$ or smaller.

Almost all sources fall in the range n(H) $= 50-200 \pccc$,
with n(H) $= 500 \pccc$ toward J0502. Similar densities in diffuse
molecular gas were inferred by \cite{NeuWel+24} from a study of
excitation in C$_2$.

\subsection{HCO$^+$ absorption and CO emission}

Figure 8 relates the \hcop\ emission brightness to the \cotw\ emission. The typical 
brightness ratios for the new line profiles on a channel-channel basis in the left panel are 
in the range 1/200-1/50, and half the aggregate \hcop\ emission occurs at \cotw\ 
brightness \WCO\ $\leq$ 2.3 K. The right panel in Figure 8 places the new results in a broader 
context by combining the new results with integrated profile results reported away from 
continuum background sources by \cite{Lis20} that extend to higher values of the 
brightnesses.  There is a wide range of \hcop\ brightness at \WCO\ $\ga 6$K-\kms\ ,
but it is within the typical range 1/200 - 1/50.

\subsection{Host of \HH\ }

The optical depth is proportional to the column density per unit velocity, and Figure 9 shows 
the \hcop\ optical depth scaled to dN(\HH)/dv and plotted against several quantities 
related to the CO emission on a channel-by-channel basis. The \cotw\ emission 
brightness (left), the CO-\HH\ conversion factor (middle; derivable from the left plot), and against the \coth/\cotw\ brightness ratio (left).

The middle panel of Figure 9 shows that most of the \HH\ 
is found at CO-\HH\ conversion factors in the range $1-4\times10^{20}\HH\pcc$ (K-\kms)$^{-1}$, but J0439 
has only very weak \cotw\ emission and a much higher ratio  
$9\times10^{20}\HH\pcc$ (K-\kms)$^{-1}$.

The right panel in Figure 9 shows that half the total amount of \HH\ occurs at 
\cotw/\coth\ brightness ratios at or above 20 and 80\% above a ratio of 10,
indicating the diffuse nature of the gas and the general absence of
saturation in \cotw. The \cotw/\coth\ ratio is expected to have been
substantially diminished by fractionation resulting from C\p\ exchange.

\begin{figure*}
\includegraphics[height=14cm]{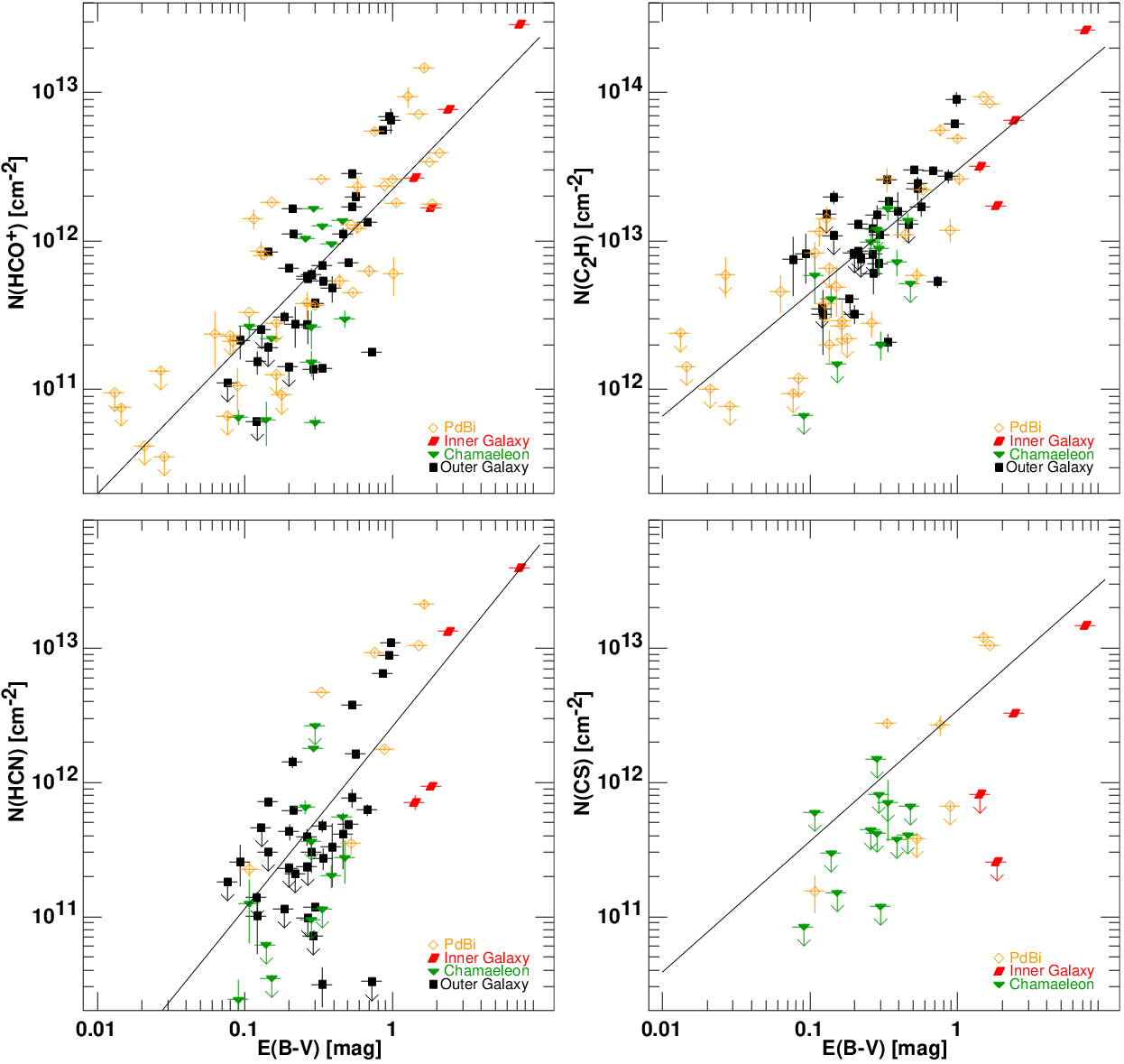}
  \caption[]{Molecular column densities plotted against \EBV\ 
from \cite{SchFin+98}. The associated datasets are indicated
(see Section 2.4). The solid line in each panel is a regression line 
for the sight lines with detections of both plotted quantities.
The values for the statistical parameters and the regression fits are given
in Table 2, and the statistical detection rates are given in Table 3.}
\end{figure*}

\begin{figure*}
\includegraphics[height=17cm]{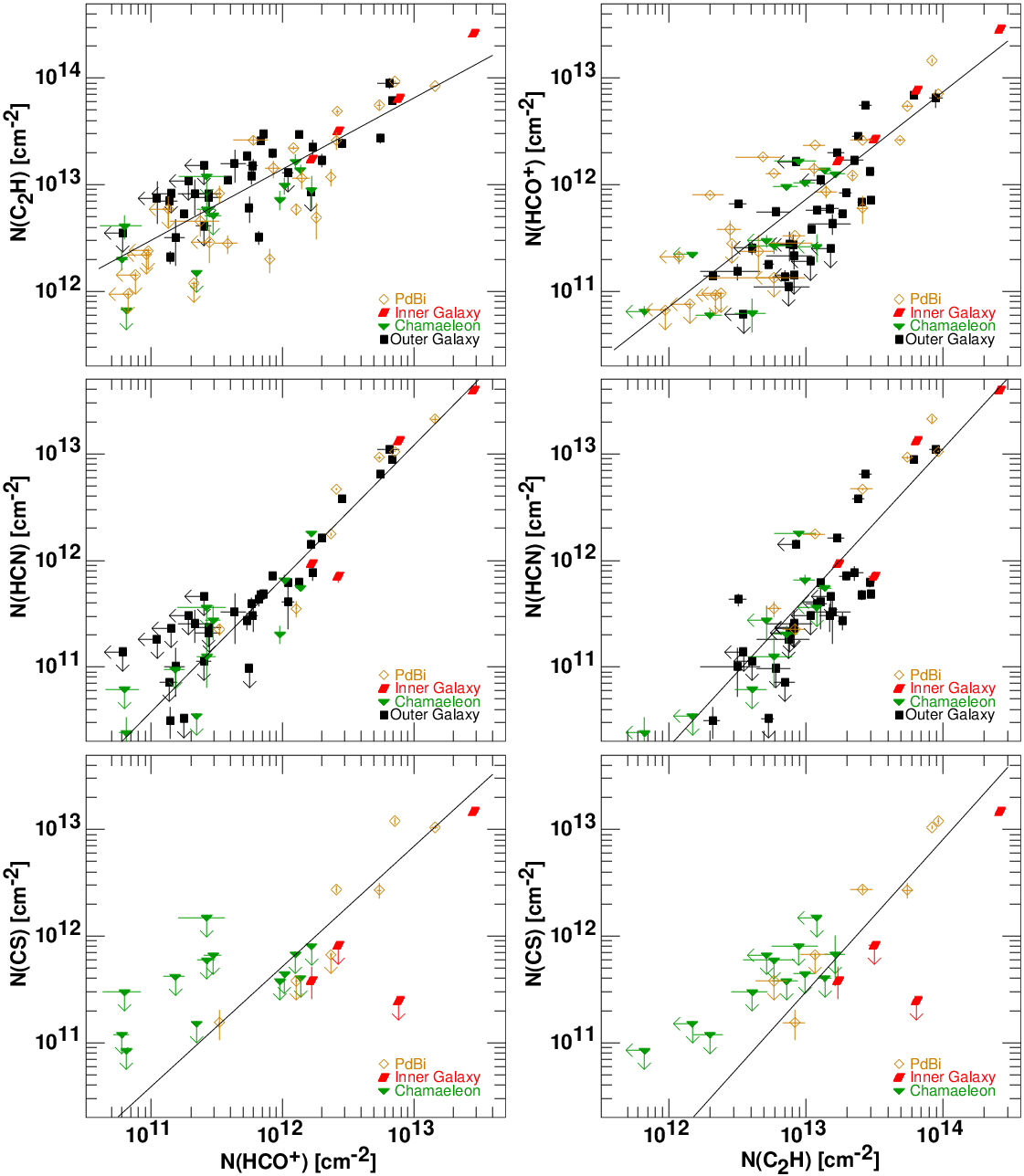}
\caption[]{
Molecular column densities plotted against N(\hcop; left) and N(\cch; right). The regression lines include only data for which both molecules were detected. 
The values for the statistical parameters and the regression fits are given in Table 5.}
\end{figure*}

\begin{figure*}
\includegraphics[height=15cm]{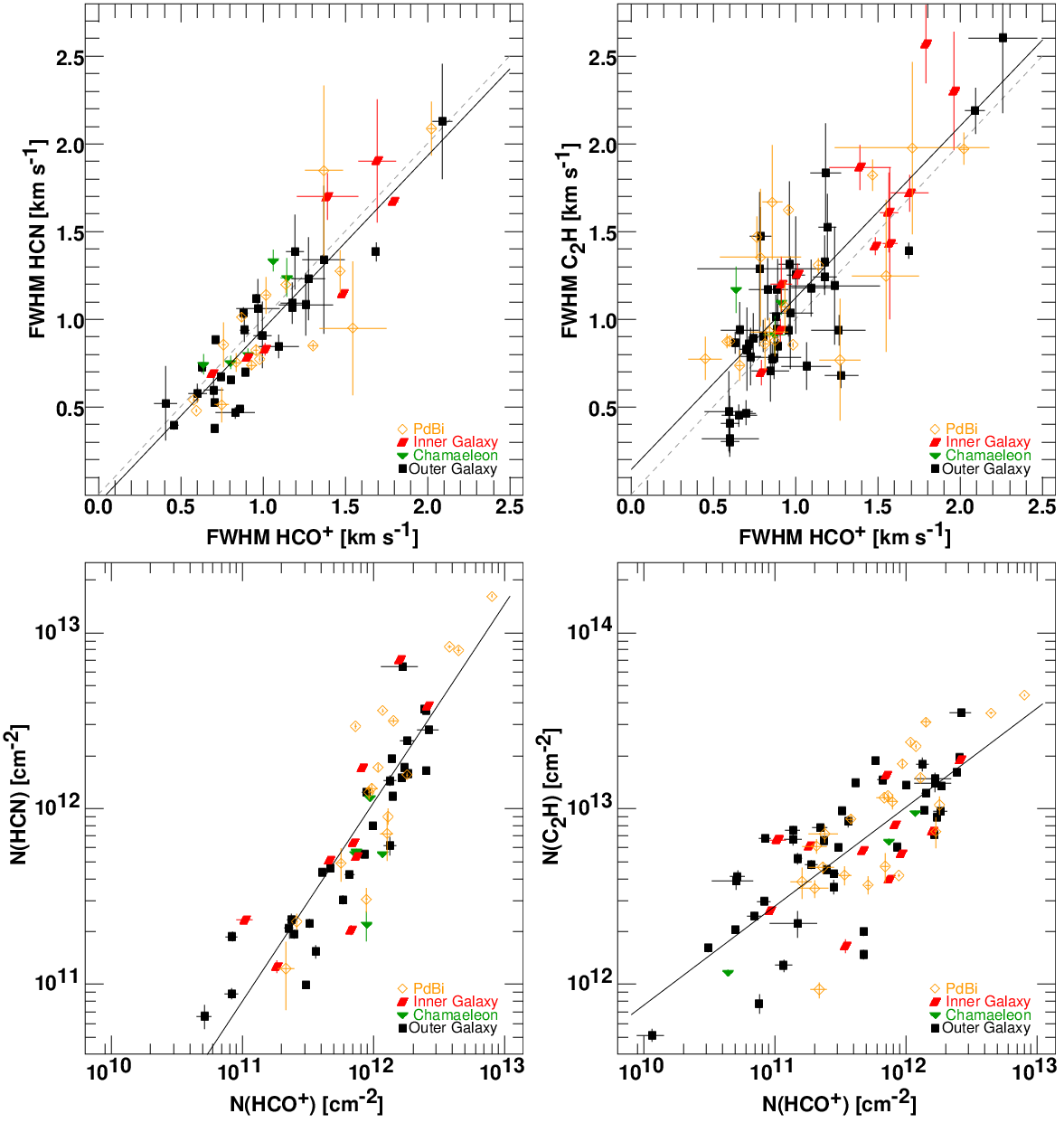}
  \caption[]{Results of the Gaussian component fitting. Top: FWHM of the Gaussian components fit to spectra of HCN and \cch\ compared 
   with those fit to \hcop. The solid black line shows the regression fit in each case, 
   and the dashed light gray line shows the locus of the equal FWHM. The numerical quantities
  derived from the fitting are given in Tables 6 and 7. Bottom: Column densities.}
\end{figure*}

\begin{figure*}
\includegraphics[height=6.9cm]{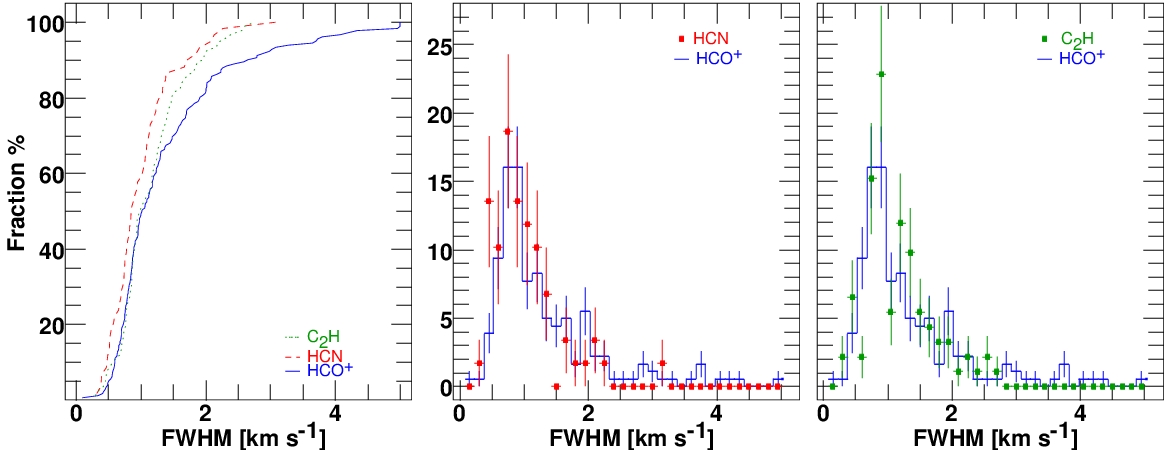}
  \caption[]{line width probability distributions for all components fit to
each species.  Left: Cumulative distributions for \hcop, HCN, and \cch. 
Middle: Fractional distributions of individual \hcop\ and HCN line widths. 
Right: Fractional distributions of \hcop\ and \cch\ line widths.}
\end{figure*}

\begin{figure*}
\includegraphics[height=8.8cm]{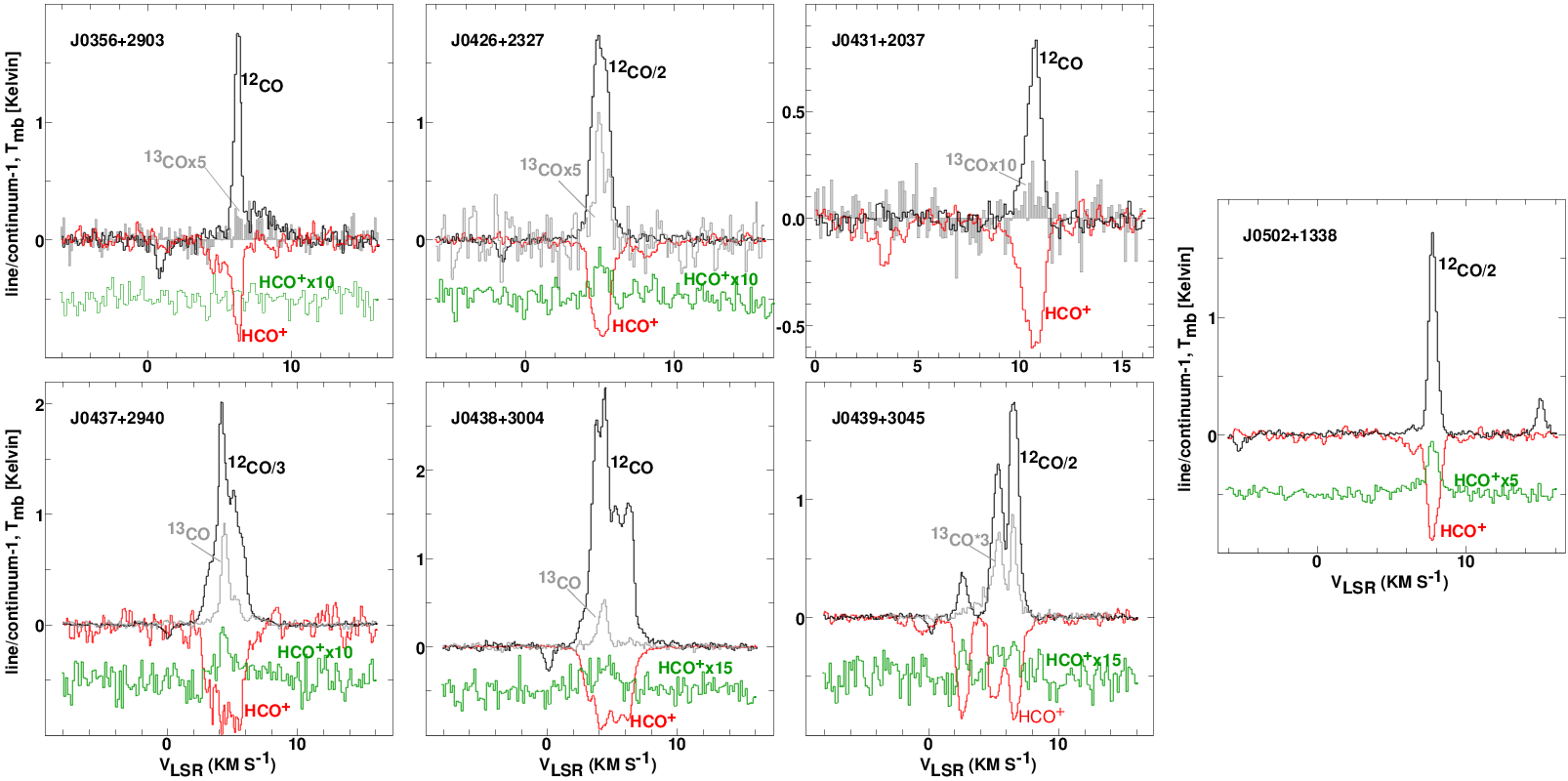}
\caption[]{
Emission profiles of \cotw (black), \coth (gray and gray shaded), \hcop (green),
and \hcop\ absorption (red). The emission profiles were normalized 
as shown.  The CO emission profiles were folded in frequency, resulting
in negative artifacts at velocities not associated with other material.}
\end{figure*}

\begin{figure*}
\includegraphics[height=7.3cm]{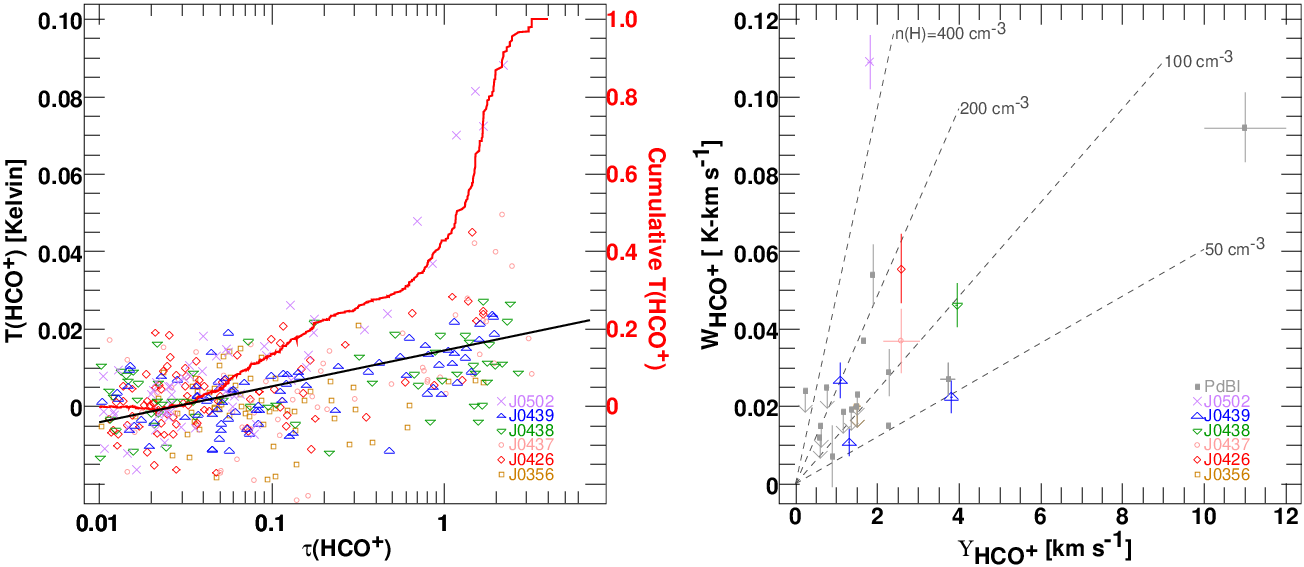}
\caption[]{
\hcop\ emission plotted against absorption.  Left: Channel values of emission 
brightness and absorption for the new data shown in Figure 6. The excitation 
temperature corresponding to the regression line is 2.758 K. 
Right: Integrated emission brightness against integrated 
absorption \Whcop\ for the new and old data.
Superposed at the right are the locii of  constant number density n(H)
for diffuse purely molecular gas using results for the weak excitation
limit (including electrons) described by \cite{LisPet16} (see Section 5.1 for an explanation of the underlying physics).}
\end{figure*}

\begin{table*}
\caption{ Statistical parameters and power-law fits for the data shown in Figure 1$^1$}
{
\small
\begin{tabular}{cccccccc}
\hline
x-species & y-species & n & $<$x$>$ & $<$y$>$ & a & b & r \\
\hline
$\sin(|b|)$ & \EBV\ & 88 & 0.348(0.226) & 0.560(0.899) mag & 1.234(0.060) & -0.999(0.083) & 0.79 \\
            & N(HI) & 88 & 0.351(0.237) & 2.137(2.878)$\times10^{21}\pcc$ & 20.592(0.033) & -0.871(0.045) & 0.90 \\
            & N(\HH) & 86 & 0.348(0.226) & 6.002(12.42)$\times10^{20}\pcc$ &19.801(0.093) & -0.903(0.130) & 0.60 \\
            & \WCO & 54 & 0.320(0.197) & 2.313(3.537) K-\kms  & -0.611(0.161) & -0.817(0.226) & 0.45 \\
\hline
\end{tabular}
\\
$^1$ The quantities in parentheses are the standard deviation. a and b are fits to log (value of the y-species) = \\
a+b$\times$log (value of the x-species), and r is the correlation coefficient.
}
\end{table*}

\begin{table*}
\caption{ Statistical parameters and power-law fits for the data shown in Figure 2}
{
\small
\begin{tabular}{cccccccc}
\hline
x-species & y-species & n & $<$x$>$ & $<$y$>$ & a & b & r \\
\hline
\EBV  & N(\hcop) & 73 & 0.656(0.948) mag & 2.130(4.011)$\times10^{12}\pcc$ & 12.347(0.065) & 1.025(0.113) & 0.733 \\
      & N(\cch) & 53 & 0.650(1.040) mag& 2.513(3.930)$\times10^{13}\pcc$ & 13.477(0.062) & 0.828(0.105) & 0.74 \\
      & N(HCN) & 38 & 0.754(1.176) mag & 3.641(7.347)$\times10^{12}\pcc$ & 12.410(0.106) & 1.355(0.197) & 0.74 \\
      & N(CS)  & 7 &2.016(2.300) mag & 6.591(5.265)$\times10^{12}\pcc$ &12.538(0.143) & 0.975(0.258) & 0.860 \\
\hline
\end{tabular}
\\
$^1$ The quantities in parentheses are the standard deviation. a and b are fits to log (value of the y-species) = \\
a+b$\times$log (value of the x-species), and r is the correlation coefficient.
}
\end{table*}

\begin{table*}
\caption{Detection rates as they vary with \EBV$^1$ and confidence level}
{
\small
\begin{tabular}{ccccccccc}
\hline
&&&&$2\sigma$ limits &&&& \\
\hline
&&&&\EBV &&&& \\
Species  & All  &0.0..0.067&0.067..0.125&0.125..0.250&0.250..0.5&0.5..1.0&1.0..2.0&2.0..7.345 \\
\hline
\hcop &  72/86&0/4   &8/12   &10/16   &24/24   &17/17   &6/6   &7/7    \\
\cch &  53/76&0/4   &6/11   &8/17   &17/21   &14/15   &4/4   &4/4    \\
HCN &  38/57&      &4/7   &4/10   &12/21   &12/13   &2/2   &4/4    \\
CS &   7/24&      &1/3   &0/2   &1/9   &1/4   &1/2   &3/4    \\
HCO &   4/46&      &0/6   &1/10   &1/20   &2/10   &      &       \\
\hline
&&&&$3\sigma$ limits &&&& \\
\hline
&&&&\EBV &&&& \\
Species  & All  &0.0..0.067&0.067..0.125&0.125..0.250&0.250..0.5&0.5..1.0&1.0..2.0&2.0..7.345 \\
\hline
\hcop &  71/86&0/4   &8/12   &9/16   &24/24   &17/17   &6/6   &7/7    \\
\cch &  45/76&0/4   &2/11   &6/17   &15/21   &14/15   &4/4   &4/4    \\
HCN &  32/57&      &1/7   &4/10   &10/21   &11/13   &2/2   &4/4    \\
CS &   7/24&      &1/3   &0/2   &1/9   &1/4   &1/2   &3/4    \\
HCO &  2/46&      &0/6   &1/10   &0/20   &1/10   &      &       \\
\hline
\end{tabular}
\\
$^1$ Entries are \#sight lines where detected/\#sight lines observed
}
\end{table*}

\begin{table}
\caption{Relative abundances of HCO}
{
\small
\begin{tabular}{lcccc}
\hline
Source & \EBV\ & N(\HH) & N(HCO) & X(HCO)$^1$ \\
       & mag   &$10^{21}\pcc$ &$10^{12}\pcc$& $10^{-9}$ \\
\hline
  J0325+2224& 0.21&0.371&0.40(0.13)& 1.08(0.34) \\
  J0433+0521& 0.30&0.127&0.17(0.06)& 1.33(0.49) \\
  J0438+3004& 0.95&2.293&2.82(0.25)& 1.23(0.11) \\
  J0439+3045& 0.87&1.850&2.16(0.82)& 1.17(0.44) \\
\hline
\bll $^2$ & 0.32 &0.867& 0.85(0.02) & 0.98(0.02) \\
\hline
\end{tabular}
\\
$^1$ X(HCO)=N(HCO)/N(\HH) \\
$^2$ Data from \cite{LisPet+14} 
\\
}
\end{table}

\begin{table*}
\caption{ Statistical parameters and power-law fits for the data shown in Figure 3}
{
\small
\begin{tabular}{cccccccc}
\hline
x-species & y-species & n & $<$x$>$ & $<$y$>$ & a & b & r \\
\hline
N(\hcop) & N(\cch) & 51 & 2.469(0.456)$\times10^{12}\pcc$ & 2.619(4.018)$\times10^{13}\pcc$ & 5.138(0.799) &0.667(0.066) & 0.82 \\
        & N(HCN) & 39 &  2.987(5.093)$\times10^{12}\pcc$ & 3.732(7.419)$\times10^{12}\pcc$ & -3.152(0.726) & 1.248(0.060) & 0.96 \\
      & N(CS)  & 7 & 8.657(9.323)$\times10^{12}\pcc$ &6.179(5.610)$\times10^{12}\pcc$ &-1.806(2.344)& 1.126(0.185) & 0.94 \\
N(\cch) & N(\hcop) &51  & 2.619(4.018)$\times10^{13}\pcc$& 2.469(0.456)$\times10^{12}\pcc$ &-1.250(1.321) & 1.008(0.100) & 0.82 \\
       & N(HCN) & 35 &  3.275(4.643)$\times10^{13}\pcc$ & 4.097(7.746)$\times10^{12}\pcc$ & -6.494(1.857) & 1.395(0.140) & 0.87 \\
       & N(CS) & 7 & 7.826(8.128)$\times10^{13}\pcc$ & 6.179(5.610)$\times10^{12}\pcc$ & -7.101(3.274) &1.429(0.239) & 0.94 \\
\hline
\end{tabular}
\\
$^1$ The quantities in parentheses are the standard deviation. a and b are fits to log (value of the y-species) = \\
a+b$\times$log (value of the x-species), and r is the correlation coefficient.
}
\end{table*}

\begin{table*}
\caption{Mean line widths and column densities of the Gaussian components 
on 28 sight lines decomposed in \hcop and HCN}
{
\small
\begin{tabular}{lccccccc}
\hline
Molecule   &  &      \hcop  & & &  HCN  & & Slope\\
\hline
Sample        & Total  &   Matched  & Unmatched   &     Total &   Matched  &  Unmatched & Matched\\
\hline
\#components    &   107    &   55  &      52  &      59  &      55  &      4 & 55 \\
$<$FWHM$>$ \kms  &   1.25(0.64)   &   1.03(0.37)  &    1.47(0.57)   &   1.03(0.52)  &    0.98(0.41)  &    1.51(1.19) & 0.99(0.07) \\
$<$N$>$ $10^{12}\pcc$   &  0.73(1.06)  &    1.26(1.25)   &   0.18(0.20)$^1$  & 1.78(2.67) & 1.88(2.74)& 0.41(0.22)$^2$ & 1.13(0.08) \\
\hline
\end{tabular}
\\
$^1$ The unmatched components of \hcop\ comprise 12\% of the total \hcop\ column density. \\
$^2$ The unmatched components of HCN comprise 2\% of the total HCN column density. \\
}
\end{table*}

\begin{table*}
\caption{Mean line widths and column densities for the Gaussian components 
on 31 sight lines decomposed in \hcop\ and \cch}
{
\small
\begin{tabular}{lccccccc}
\hline
Molecule    &  &   \hcop  &  &   &   \cch  &   \\
\hline
Sample  &   Total  &   Matched  &  Unmatched  &   Total   &  Matched  &  Unmatched  & Slope\\
\hline
\#components   &    119   &    73   &     46   &     81  &      73  &      8  & 73 \\
$<$FWHM$>$ \kms  &  1.26(0.73)  &   1.03(0.37)   &  1.66(0.90)  & 1.20(0.56) & 1.15(0.49)& 1.48(0.82) & 0.98(0.09) \\
$<$N$>$ $10^{12}\pcc$  &  0.71(1.00)   &  0.92(1.17)  &   0.20(0.51)$^1$& 9.29(8.28)& 9.76(0.81) &5.00(3.10)$^2$ & 0.56(0.06) \\
\hline
\end{tabular}
\\
$^1$ The unmatched components of \hcop\ comprise 11\% of the total \hcop\ column density. \\
$^2$ The unmatched components of \cch\ comprise 5\% of the total \cch\ column density. \\
}
\end{table*}

\section{Summary}
We joined recent ALMA observations (Table 1) with older absorption data taken 
at the IRAM Plateau de Bure Interferometer (PdBI) and Northern Extended Millimeter Array (NOEMA) telescopes to analyze a larger body of absorption spectra 
of \hcop, \cch, HCN, CS, and HCO arising in diffuse interstellar gas seen toward compact extragalactic continuum sources.  
The sight lines in the sample are slightly overabundant on average in 
\EBV\ compared to the Galactic average (Figure 1, Section 3.1) and richer in 
reddening than in the column density of atomic hydrogen. This implies that the sample
is not only rich in gas and dust, but has a higher fraction of \HH. The distribution of \EBV\ with $\sin(|b|)$ has a higher scatter than that 
of N(HI), which can be understood given the large scatter in the plots of N(\HH) and integrated CO emission \WCO . 

Figure 2  plots the column densities of \hcop, \cch, HCN, and CS against \EBV\ 
(see Section 3.2 and Table 2). Absorption from \hcop, the most 
ubiquitous species (72 of 86 sight lines; Table 3), was 
uniformly detected at \EBV\ $\ga 0.2$ mag and was detected more rarely down to 
\EBV\ = 0.07 mag, where \HH\ first becomes a substantial component of the ISM. 
\cch\ is ubiquitous, and N(\cch)/N(\hcop) increases at smaller N(\hcop), but 
\cch\ was detected slightly less frequently than \hcop\ (53 of 76 sight lines) 
because \cch\ absorption is  intrinsically weaker and harder to detect.
The so-called dense-gas tracer HCN was uniformly detected down to \EBV\ = 0.3 
mag and occasionally seen even below \EBV\ = 0.1 mag.  It was uniformly
detected at N(\HH) $ \ga 10^{20}\pcc$ with little change in the relative abundance 
N(HCN)/N(\hcop) $\approx 1.25$ or N(HCN)/N(\HH) $= 3.75\times10^{-9}$.  

We presented no new detections of CS, which we failed to detect along all 13 
sight lines we studied in the outskirts of Chamaeleon (owing to a lack of sensitivity),
and which we subsequently did not observe in the Galactic anticenter.  The upper
limits achieved in the Chamaeleon study show, however, that the relative abundance 
of CS is lower at the low-\EBV\ end of our sample at \EBV\ $\la 1$ mag (Fig. 2) 
and perhaps (Fig. 3) when N(\hcop) $\la 2\times10^{12}\pcc$.  

HCO is ubiquitous 
in the diffuse ISM with N(HCO)/N(\hcop) = 1/3 or N(HCO)/N(\HH) $= 1\times10^{-9}$ ,
but the strongest hyperfine component of HCO is only about 2\% as strong in 
absorption as \hcop\ and was only detected along the four ALMA sight lines 
with a signal-to-noise ratio of 100:1 or more in \hcop\ (Table 5).   

Figure 3 (see Section 3.3 and Table 5) shows the inter-relations among the column densities of \hcop, \cch, HCN, and CS.  
HCN is detected almost without exception down to 
N(\hcop) $= 3\times 10^{11}\pcc$ or N(\HH) $= 10^{20}\pcc$, in contrast to 
earlier results that implied that N(HCN)/N(\hcop) increased abruptly at N(\hcop) $= 10^{12}\pcc$.
The bottom panel in Figure 3 shows that the lack of detections of CS at \EBV\ $<$ 0.5 mag might result from a lack of sensitivity, because the observed upper limits on N(CS) are higher than the values expected from extrapolating the detections
at higher N(\hcop) and N(\cch).

\begin{figure*}
\includegraphics[height=7.3cm]{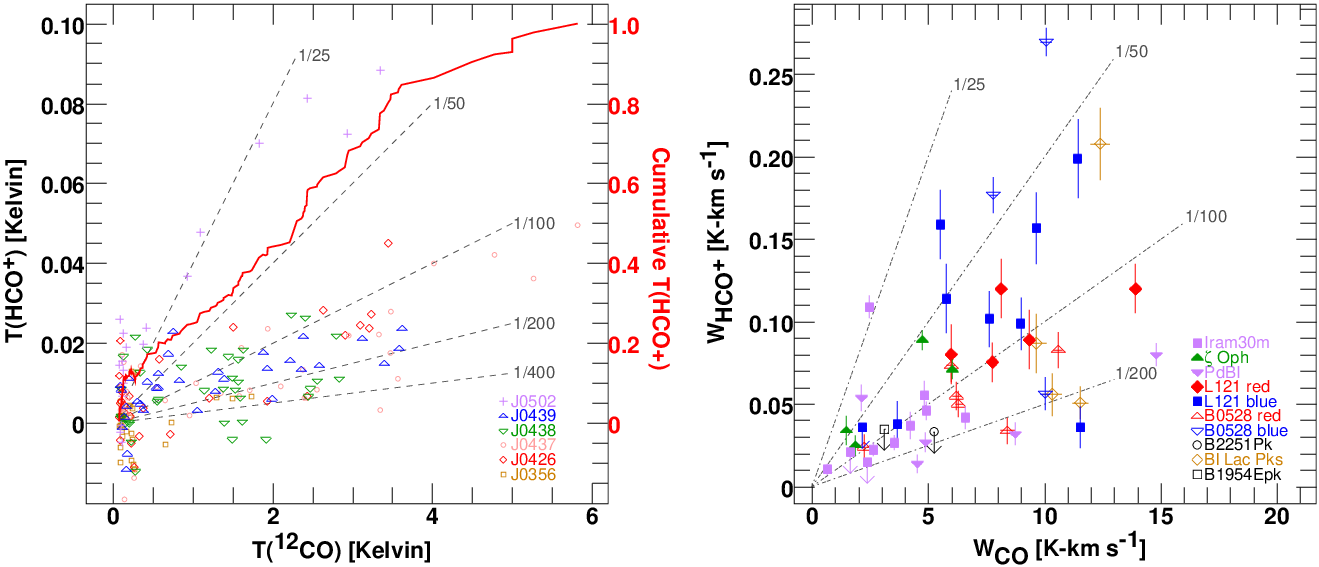}
\caption[]{
Comparison of emission brightness for \hcop\ and \cotw. Left: Channel-channel
comparison for the six new anticenter sources with \WCO\ $>$ 1 K-\kms.  Right: 
Integrated \hcop\ emission plotted against \WCO\ with new data from this work
overlaid on earlier observations  taken around and well removed from continuum 
background sources. This was previously shown as Figure 8 in \cite{Lis20}.}
\end{figure*}

In Section 4 we discussed the Gaussian decomposition of ALMA 
profiles of \hcop, \cch, and HCN by comparing the numbers and line widths 
of kinematic components in Figures 4-5 and Tables 6 and 7. The typical line widths for all species are 1 \kms.  \hcop\ 
is the most widely detected species as a result of its high dipole moment 
and lack of hyperfine structure, and its spectra are richer and decomposed into more components than in \cch\ or HCN.  For HCN (Table 7), 55 of 59 HCN components seen along 28 sight lines were matched against 55 of 107 components in \hcop, and the matched HCN components were found to be 5\% narrower on average (0.975 versus 1.032 \kms).  The unmatched components are weaker and broader than average and constitute small fractions of the total absorption in either species. For \cch\ (Figures 4-5 and Table 7) 73 of  81 components were matched against 73 of 119 \hcop\ components along 31 sight lines,
and the matched \cch\ components are 12\% broader, 1.12 versus 1.03 \kms.
The unmatched components constitute a small fraction of the total
absorption.

\begin{figure*}
\includegraphics[height=5.8cm]{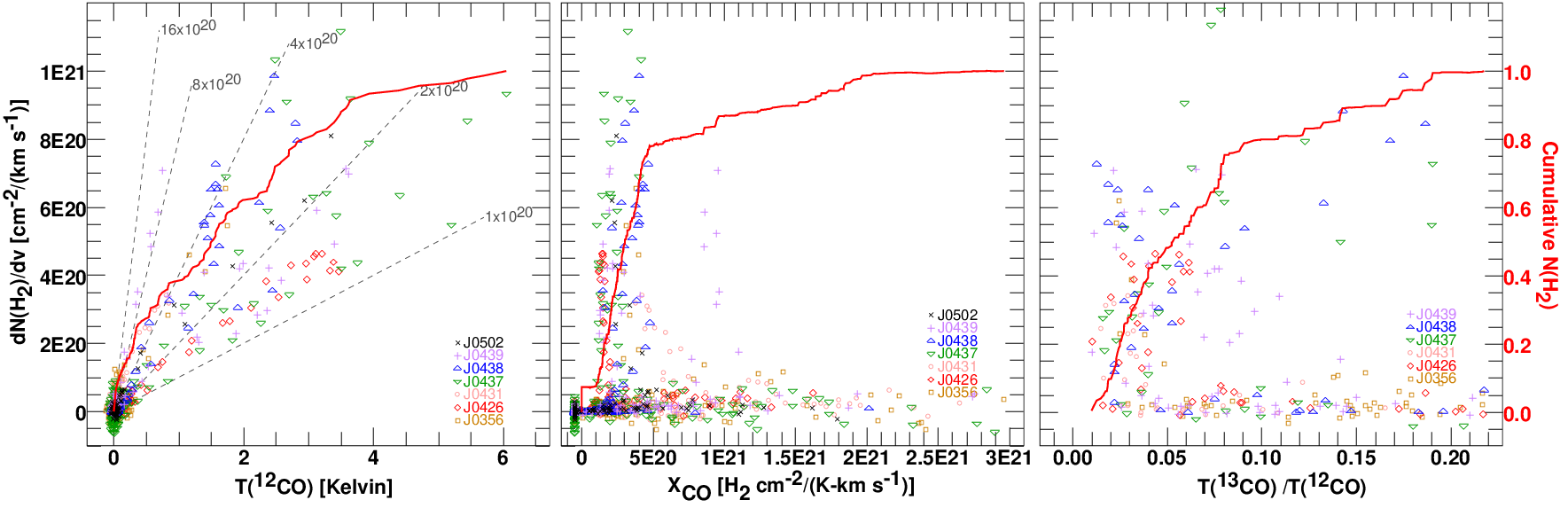}
\caption[]{
Channel-by-channel comparison with values of the \hcop\ absorption scaled to \HH. 
The fat red curve in each panel is the fractional cumulative distribution, which is to 
be read against the red scale at the right.}
\end{figure*}

In Section 5 we discussed new observations of \hcop\ and CO emission along seven outer Galaxy sight lines (Figure 6). 
A channel-by-channel comparison of the \hcop\ emission and absorption 
showed (Figure 7, left) that \hcop\ emission is detected 
in a statistical sense at \hcop\ optical depths as low as 0
.05, and the great majority of channels imply a typical excitation temperature of 2.76 K up to $\tau$ = 2.  
Stronger emission with an excitation temperature 2.86 K was seen
on the the sight line to J0502. Deriving N(\hcop) in the limit of no collisional excitation is a good approximation.

The comparison of the integrated \hcop\ emission and absorption (Figure 7, right) showed that the mean hydrogen volume density falls in the range 
n(H) $= 50 - 200\pccc$ on six of the seven sight lines, with n(H) $= 500\pccc$ toward J0502, where the \hcop\ emission is brighter.

The brightness ratio of the \hcop\ emission to that of CO is generally within a factor of two of 1\% for gas seen near the continuum background targets (Figure 8, left).  Most of the \hcop\ emission arises in channels in which the CO brightness is 2 K or lower.

In Section 5.1 (Figure 9) we related the \HH\ content to the nature of the host gas as characterized by different aspects of the CO emission. Nearly half of the total amount of \HH\ arises in line channels with a CO brightness of $\la 2$ K (Figure 9, left) and about 80\% with CO-\HH\ conversion factors N(\HH)/\WCO\ = $1-5 \times 10^{20}$ \HH\ (K-\kms)$^{-1}$ and \cotw/\coth\ emission brightness ratios
above 10.
 
\begin{acknowledgements}

The National Radio Astronomy Observatory is a facility of the
National Science Foundation, operated by Associated Universities, Inc.

This work was supported by the French program “Physique et Chimie du Milieu 
Interstellaire” (PCMI) funded by the Conseil National de la Recherche Scientifique 
(CNRS) and Centre National d'Etudes Spatiales (CNES).

This work is based in part on observations carried out under project number 
003-19 with the IRAM 30m telescope]. IRAM is supported by INSU/CNRS (France), 
MPG (Germany) and IGN (Spain).

This paper makes use of the following ALMA data
\begin{itemize}
\item{ADS/JAO.ALMA\#2016.1.00132.S}
\item{ADS/JAO.ALMA\#2016.1.00714.S}
\item{ADS/JAO.ALMA\#2018.1.00115.S}
\item{ADS/JAO.ALMA\#2019.1.00120.S}
\end{itemize}

ALMA is a partnership of ESO (representing its member states), NSF (USA)
and NINS (Japan), together with NRC (Canada), NSC and ASIAA (Taiwan), and
KASI (Republic of Korea), in cooperation with the Republic of Chile.  The
Joint ALMA Observatory is operated by ESO, AUI/NRAO and NAOJ. 

We thank the referee for a close reading that resulted in a significantly
improved presentation.

\end{acknowledgements}

\bibliographystyle{apj}

\begin{appendix}

\section{Data sources}

The new and old data sources are described in Section 2.1 and Table \ref{TableDataSources}.

\begin{table}
\caption{Data sources}
{
\small
\begin{tabular}{lcccccc}
\hline
Observation   & old  & new \\
\hline
Absorption  & PdBI, Noema & ALMA \\
Emission    & ARO 12m & IRAM 30m \\
\hline
\label{TableDataSources}
\end{tabular}
}
\end{table}

\section{Line of sight properties: \hcop, HCN and \cch}

New results for \hcop, HCN and \cch\ are summarized in Table \ref{TableLOS}.

\begin{table*}
\caption{Line of sight properties, \hcop\ column densities and new results for HCN and \cch}
{
\small
\begin{tabular}{lcccccclll}
\hline
Source& RA(J2000)&Dec(J2000)& $l$ & $b$ &\EBV& flux &N(\hcop)&N(HCN) & N(\cch) \\
      &  hh.mmsss &dd.mmsss&degrees & degrees &mag &Jy&$10^{12}\pcc$&$10^{12}\pcc$ & $10^{13}\pcc$ \\  
\hline
J0203+1134& 2.03464& 11.34492&149.6826&-47.4992&0.144&0.128&$\leq$0.289&$\leq$0.455&$\leq$1.635\\
J0209+1352& 2.09357& 13.52045&150.1800&-44.8290&0.094&0.225&0.215(0.055)&$\leq$0.263&$\leq$0.908\\
J0211+1051& 2.11132& 10.51348&152.5781&-47.3674&0.145&0.465&0.841(0.032)&0.713(0.053)&1.971(0.201)\\
J0213+1820& 2.13105& 18.20255&148.7289&-40.4014&0.130&0.094&$\leq$0.379&$\leq$0.688&$\leq$2.278\\
J0231+1322& 2.31459& 13.22547&157.0917&-42.7380&0.121&0.431&0.154(0.028)&$\leq$0.148&$\leq$0.445\\
J0242+1742& 2.42243& 17.42589&157.0180&-37.7033&0.077&0.227&$\leq$0.166&$\leq$0.271&$\leq$0.967\\
J0252+1718& 2.52077& 17.18427&159.7420&-36.7885&0.220&0.205&0.276(0.084)&$\leq$0.312&$\leq$1.139\\
J0325+2224& 3.25368& 22.24004&163.6700&-28.0213&0.213&1.171&1.114(0.019)&0.620(0.031)&1.293(0.058)\\
J0329+3510& 3.29154& 35.10060&155.9152&-17.4042&0.267&0.582&0.559(0.035)&$\leq$0.146&0.606(0.169)\\
J0329+2756& 3.29577& 27.56155&160.7030&-23.0743&0.198&0.161&$\leq$0.213&$\leq$0.345&$\leq$1.240\\
J0334+0800& 3.34533&  8.00144&177.2396&-37.0871&0.391&0.151&0.483(0.097)&$\leq$0.492&$\leq$1.627\\
J0336+3218& 3.36301& 32.18293&158.9998&-18.7650&0.733&1.674&0.178(0.010)&$\leq$0.049&0.534(0.049)\\
J0356+2903& 3.56085& 29.03423&164.6120&-18.4927&0.212&0.140&1.648(0.099)&1.415(0.145)&$\leq$1.279\\
J0357+2319& 3.57216& 23.19538&169.0302&-22.4661&0.185&0.227&0.307(0.030)&$\leq$0.171&$\leq$0.612\\
J0400+0550& 4.00117&  5.50431&184.2710&-33.7266&0.266&0.161&0.273(0.070)&$\leq$0.352&$\leq$1.224\\
J0401+0413& 4.01199&  4.13344&186.0261&-34.4947&0.341&0.409&0.536(0.023)&0.272(0.051)&1.853(0.126)\\
J0403+2600& 4.03056& 26.00015&168.0260&-19.6483&0.201&0.335&0.657(0.032)&0.431(0.058)&0.322(0.047)\\
J0406+0637& 4.06343&  6.37150&184.7075&-32.0009&0.283&0.267&0.595(0.057)&0.302(0.089)&1.503(0.235)\\
J0407+0742& 4.07291&  7.42075&183.8723&-31.1558&0.265&0.390&0.580(0.034)&0.394(0.057)&1.207(0.214)\\
J0426+0518& 4.26366&  5.18199&189.3631&-28.7705&0.291&0.518&0.137(0.022)&$\leq$0.108&0.708(0.131)\\
J0426+2327& 4.26557& 23.27396&173.8881&-17.4457&0.539&0.309&2.831(0.063)&3.780(0.086)&2.434(0.249)\\
J0427+0457& 4.27476&  4.57083&189.8857&-28.7306&0.335&0.413&0.686(0.026)&0.474(0.052)&2.583(0.164)\\
J0437+2037& 4.31038& 20.37343&176.8096&-18.5565&0.532&0.221&1.696(0.081)&0.769(0.121)&2.260(0.399)\\
J0431+1731& 4.31574& 17.31358&179.4942&-20.3579&0.464&0.106&1.109(0.121)&$\leq$0.553&$\leq$1.939\\
J0433+0521& 4.33111&  5.21156&190.3730&-27.3967&0.298&4.931&0.382(0.003)&0.118(0.006)&1.099(0.019)\\
J0437+2940& 4.37044& 29.40138&170.5818&-11.6609&0.979&0.060&6.517(1.252)&10.987(0.753)&8.980(1.016)\\
J0438+3004& 4.38049& 30.04455&170.4116&-11.2283&0.952&0.692&6.878(0.042)&8.815(0.049)&6.154(0.134)\\
J0439+3045& 4.39178& 30.45076&170.0655&-10.5913&0.867&0.196&5.551(0.090)&6.438(0.110)&2.734(0.292)\\
J0440+1437& 4.40211& 14.37570&183.2538&-20.5438&0.681&0.339&1.333(0.034)&0.621(0.060)&2.956(0.202)\\
J0445+0715& 4.45014&  7.15539&190.4535&-23.8898&0.121&0.277&$\leq$0.091&$\leq$0.208&$\leq$0.526\\
J0449+1121& 4.49077& 11.21286&187.4274&-20.7365&0.504&0.526&0.714(0.024)&0.484(0.037)&3.001(0.118)\\
J0502+1338& 5.02332& 13.38110&187.4143&-16.7456&0.564&0.273&1.986(0.065)&1.626(0.075)&1.694(0.239)\\
J0510+1800& 5.10024& 18.00416&184.7304&-12.7895&0.328&2.442&0.139(0.006)&0.031(0.010)&0.209(0.029)\\
J0942-7731& 9.42427&-77.31116&293.3194&-18.3299&0.336&0.204&1.256(0.074)&$\leq$0.170&1.653(0.284)\\
J1058-8003&10.58433&-80.03542&298.0095&-18.2881&0.152&1.188&0.221(0.010)&$\leq$0.052&$\leq$0.223\\
J1136-6827&11.36021&-68.27061&296.0710& -6.5902&0.460&0.470&1.365(0.039)&0.554(0.015)&1.376(0.158)\\
J1145-6954&11.45536&-69.54018&297.3172& -7.7465&0.387&0.541&0.957(0.034)&0.203(0.039)&0.724(0.147)\\
J1147-7935&11.47334&-67.53418&296.9588& -5.7671&0.300&1.563&0.060(0.006)&$\leq$3.969&0.200(0.043)\\
J1152-8344&11.52532&-83.44094&301.2388&-21.0576&0.283&0.114&0.264(0.077)&$\leq$0.544&$\leq$1.804\\
J1224-8313&12.24544&-83.13101&302.0983&-20.3907&0.257&0.301&1.040(0.048)&0.652(0.076)&0.988(0.179)\\
J1254-7138&12.54598&-71.38184&303.2156& -8.7687&0.282&0.383&0.153(0.030)&$\leq$0.143&$\leq$0.000\\
J1312-7724&13.12387&-77.24131&304.1251&-14.5810&0.476&0.209&0.298(0.040)&$\leq$0.294&$\leq$0.777\\
J1550-8258&15.50592&-82.58065&308.2800&-22.0469&0.107&0.357&0.266(0.044)&$\leq$0.188&$\leq$0.629\\
J1617-7717&16.17493&-77.17185&313.4306&-18.8538&0.091&2.478&0.065(0.008)&$\leq$0.029&$\leq$0.100\\
J1723-7713&17.23509&-77.13502&315.6923&-21.7985&0.255&0.312&1.658(0.057)&1.799(0.057)&$\leq$0.901\\
J1733-7935&17.33407&-79.35554&313.6117&-23.2669&0.139&0.606&0.066(0.020)&$\leq$0.093&0.409(0.104)\\
\hline
\end{tabular}
\\
}
\label{TableLOS}
\end{table*}

\section{Spectra of C$_2$H}

\begin{figure*}
\includegraphics[height=18cm]{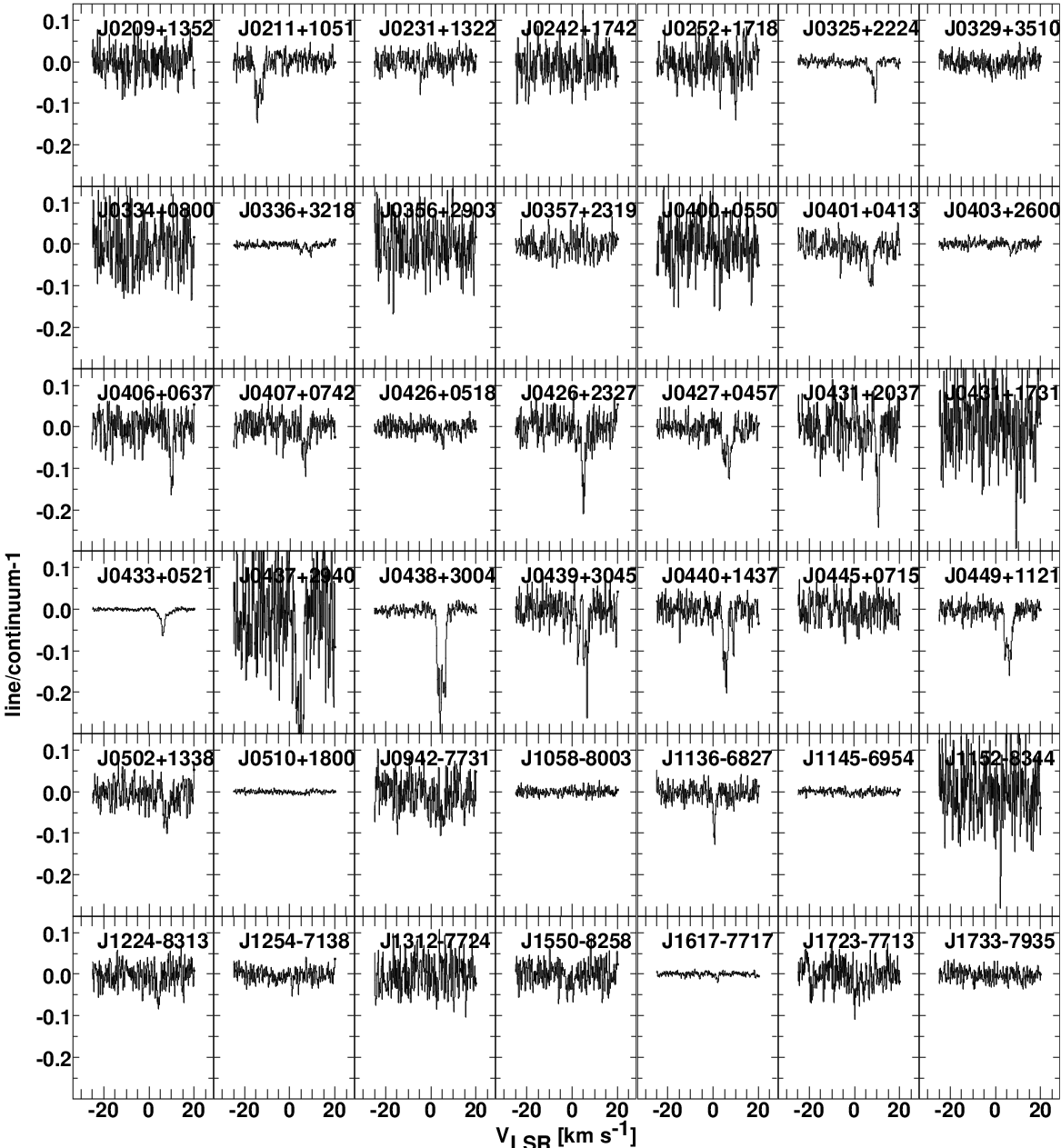}
\caption[]{Spectra of \cch\ absorption. Spectra along 4 sight lines lacking apparent detections are not shown.
See Table 1 for quantitative results and Table 4 for the statistics of detection rates.}
\label{FigureC2H}
\end{figure*}

Recent ALMA absorption spectra of \cch\ are shown in Figure \ref{FigureC2H}.

\section{Spectra of HCN}

\begin{figure*}
\includegraphics[height=14cm]{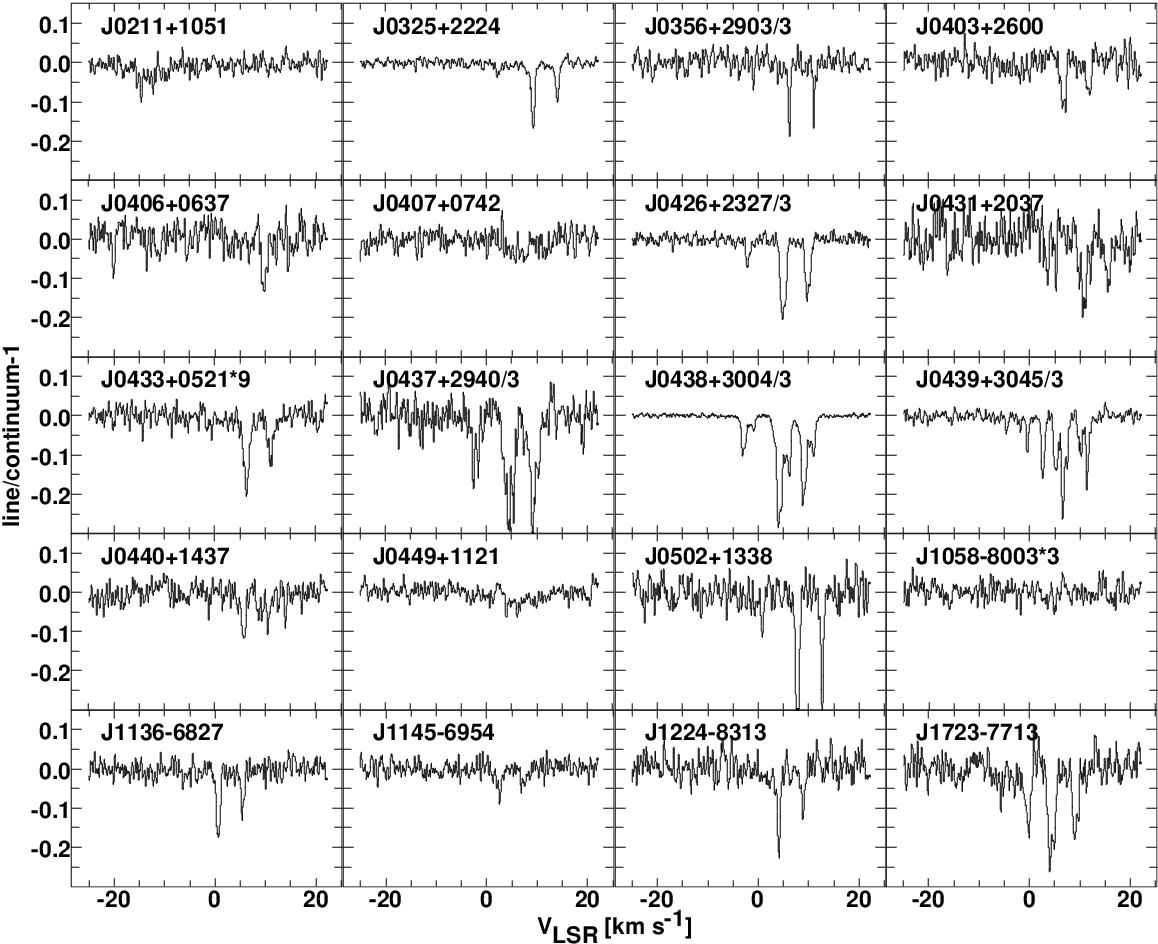}
 \caption[]{ALMA spectra of HCN absorption including all statistically significant detections. Some spectra have been scaled as indicated in those panels. See Table 1 for quantitative results
 and Table 4 for the statistics of detection rates.}
\label{FigureHCN}
\end{figure*}

Recent ALMA absorption spectra of HCN are shown in Figure \ref{FigureHCN}.

\section{Spectra of HCO}

\begin{figure*}
\includegraphics[height=15cm]{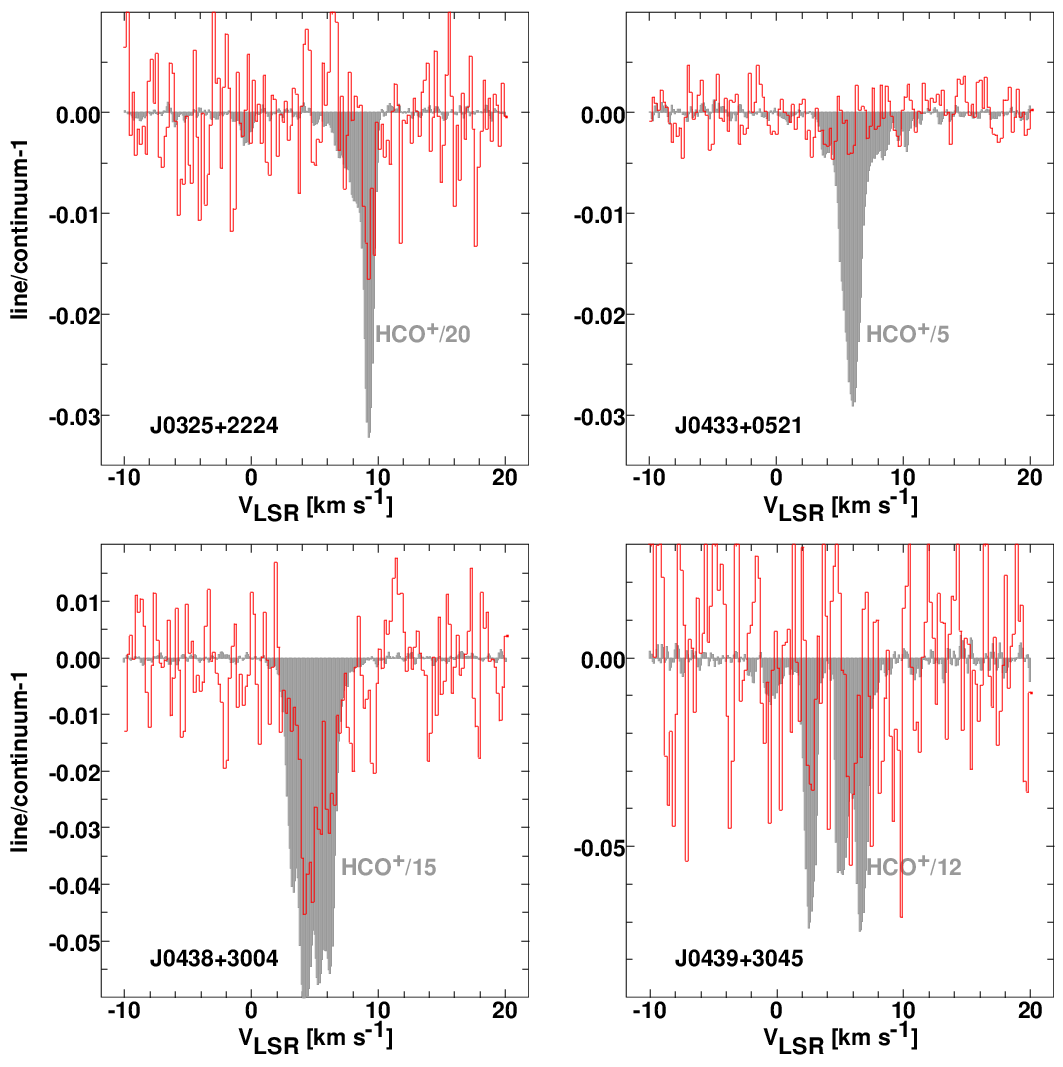}
 \caption[]{ALMA spectra of HCO absorption shown as red
 histograms superposed on scaled profiles of \hcop\ absorption shown
 as gray histograms.}
\label{FigureHCO}
\end{figure*}

 Recent ALMA absorption spectra of HCO are shown in Figure \ref{FigureHCO} for the sight lines noted in Table 5.

\section{Forming emission spectra in the presence of a continuum background.}
\label{SectionA.5}

In Section 5 we discussed the excitation of \hcop\ based on 
comparison of emission and absorption spectra of \hcop\ and CO. 
Because the emitted and absorbed fluxes of \hcop\ lines are by
coincidence comparable at most millimeter-wave telescopes, pure emission
profiles cannot be obtained directly toward the continuum \cite{LucLis96}.  Instead
we observed at positions displaced from the continuum by 1.2 HPBW = 30\arcsec\ in each of the four cardinal directions, similar to \cite{LucLis96}. Spectra around J0426+2327 are shown in Figure A.4. The individual \hcop\ spectra at right are noisy but see the average in Figure 6.

\begin{figure*}
\includegraphics[height=9cm]{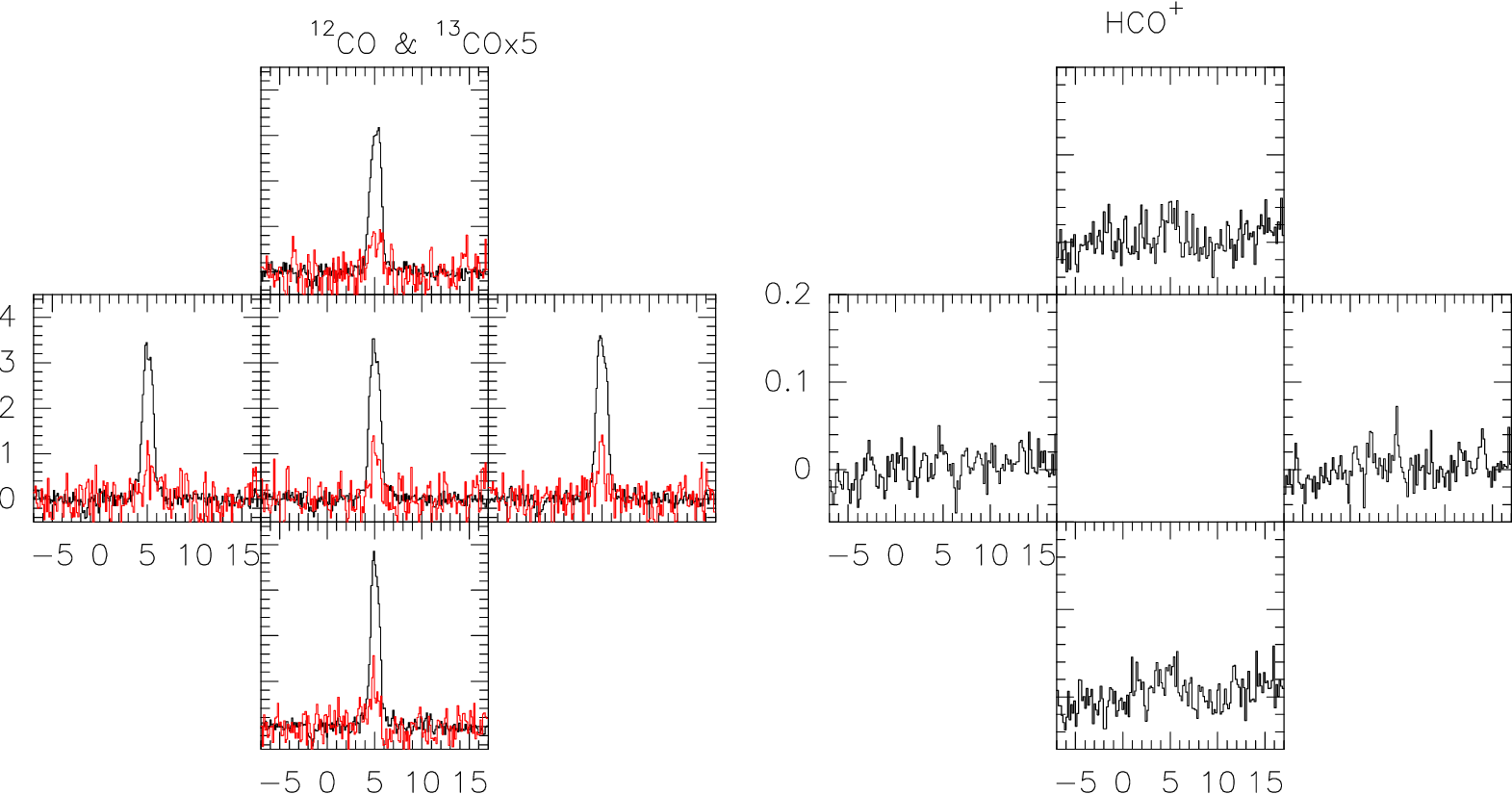}
 \caption[]{Individual spectra used to form the average spectra toward J0426+2327 as discussed in Section 5 and shown in Figure 7.  \cotw\ (black) and \coth\ spectra shown in red and scaled upward by factor 5 are shown at left and \hcop\ spectra  are shown at right. }
\label{FigureIndividualLOS}
\end{figure*}

Mixing of emission and absorption is not an issue for the much stronger CO emission lines that are shown at left in Figure \ref{FigureIndividualLOS}. The spectra toward and displaced 1 HPBW = 22\arcsec\ from the continuum differ little so we averaged all five positions to form the \cotw\ and \coth\ spectra shown in Figure 6. 

Although the averaged CO and \hcop\ spectra discussed in Section 5 were derived over slightly regions they (as earlier) seem to present a coherent picture of the  \hcop\ excitation.

\end{appendix}

\end{document}